\documentclass[a4paper,fleqn,usenatbib]{mnras}

\usepackage{newtxtext,newtxmath}
\usepackage[T1]{fontenc}
\usepackage{ae,aecompl}

\usepackage{amsmath}
\usepackage{amssymb}
\usepackage{color}
\usepackage{epic}
\usepackage{epstopdf}
\usepackage{graphicx}
\usepackage{hyperref}
\usepackage{multirow}
\usepackage{natbib}
\usepackage{overpic}

\title[RA\MakeLowercase{i}SE {II}: resolved spectral evolution in radio AGN]{RA\MakeLowercase{i}SE {II}: resolved spectral evolution in radio AGN}

\author[R. J. Turner et al.]{
Ross J. Turner$^{1,2}$\thanks{Email: turner.rj@icloud.com},
Jonathan G. Rogers$^{1,2}$,
Stanislav S. Shabala$^{1}$ and\newauthor
\:\,Martin G. H. Krause$^{1,3}$\\
$^{1}$School of Physical Sciences, University of Tasmania, Private Bag 37, Hobart, 7001, Australia\\
$^{2}$CSIRO Astronomy and Space Science, Post Office Box 76, Epping, New South Wales 1710, Australia\\
$^{3}$Centre for Astrophysics Research, School of Physics, Astronomy and Mathematics, University of Hertfordshire, College Lane,\\ Hatfield AL10 9AB, UK}

\date{Accepted 2017 October 03. Received 2017 October 03; in original form 2017 July 21}

\pubyear{2017}

\begin{document}
\label{firstpage}
\pagerange{\pageref{firstpage}--\pageref{lastpage}}
\maketitle

\begin{abstract}

{The active galactic nuclei (AGN) lobe radio luminosities modelled in hydrodynamical simulations and most analytical models do not address the redistribution of the electron energies due to adiabatic expansion, synchrotron radiation and inverse-Compton scattering of CMB photons. We present a synchrotron emissivity model for resolved sources which includes a full treatment of the loss mechanisms spatially across the lobe, and apply it to a dynamical radio source model with known pressure and volume expansion rates. The bulk flow and dispersion of discrete electron packets is represented by tracer fields in hydrodynamical simulations; we show that the mixing of different aged electrons strongly effects the spectrum at each point of the radio map in high-powered FR-II sources. The inclusion of this mixing leads to a factor of a few discrepancy between the spectral age measured using impulsive injection models (e.g. JP model) and the dynamical age. The observable properties of radio sources are predicted to be strongly frequency dependent: FR-II lobes are expected to appear more elongated at higher frequencies, while jetted FR-I sources appear less extended. 
The emerging FR0 class of radio sources, comprising gigahertz peaked and compact steep spectrum sources, can potentially be explained by a population of low-powered FR-Is. The extended emission from such sources is shown to be undetectable for objects within a few orders of magnitude of the survey detection limit and to not contribute to the curvature of the radio SED.}

\end{abstract}


\section{INTRODUCTION}
\label{sec:INTRODUCTION}

This is the {second} in a series of papers presenting an analytical model for the evolution of active galactic nuclei (AGN) radio sources. In our first paper \citep{Turner+2015} we developed a model, \emph{Radio AGN in Semi-analytic Environments} (RAiSE), for the dynamical evolution of the lobes of AGNs; {it builds on the theoretical framework developed in previous models \citep[e.g.][]{KA+1997, Blundell+1999, KC+2002, Luo+2010}. Our model relaxes several restrictions imposed in this earlier work including on the profile of the host halo environment and the self-similar expansion of the pressure-inflated lobe. Some simplifying assumptions are retained to keep the problem analytically tractable, e.g. homogeneous lobe magnetic fields.} 
{In a companion paper} \citep{Turner+2016}, we develop robust Bayesian algorithms for fitting the jet powers, ages and magnetic field strengths of radio sources, including in situations when incomplete observational data are available.
Here we provide a general analytic form for the radio emission.

Radio-loud AGNs consist of twin jets emanating in opposite directions from the regions around the supermassive black hole, typically located near the centre of its host galaxy. These jets interact with the surrounding gas, inflating an underdense ``lobe'' with synchrotron radio emission arising from shocked leptons injected by the jet. These radio sources are separated into two classes based on the observed distribution of radio surface brightness in their lobes \citep{FR+1974}. \citeauthor{FR+1974} type II sources have well-defined jet termination shocks, known as ``hotspots'', located towards the ends of edge-brightened lobes; type I sources have their regions of highest surface brightness close to the core. The \citet{Turner+2015} model simulates the evolution of radio sources which commence their expansion with an FR-II type morphology, though may later transition into pressure-limited expansion for weaker sources. The behaviour of the jet and the injected synchrotron-emitting electrons largely determine the observed morphological class. The jet plasma generates a recollimation shock when the sidewards ram pressure of the conical jet matches the pressure of the ambient medium \citep{Krause+2012}. This recollimation shock stabilises the jet if it reaches the jet axis before the jet runs out of thrust, leading to an FR-II morphology. However, if the jet cannot be stabilised a flare point as seen in observations of type I objects forms. The outward flow of plasma may lead to backflow and the inflation of radio lobes, or the prototypical smoke-like plumes \citep[e.g. 3C31;][]{Leahy+2013}; a different dynamical model is required for such ``flaring jet'' sources.

In this paper we extend the dynamical model detailed in \citet{Turner+2015} for initially FR-II type radio sources to include an analytic prescription for the spatial distribution of the synchrotron lobe luminosity. In Section \ref{sec:LUMINOSITY MODEL} we describe a general luminosity model which is applicable for radio sources expanding into environments consistent with literature X-ray observations, and with arbitrary volume and pressure evolution (e.g. not self-similar as is commonly assumed in dynamical models). We then discuss an analytical model for the three-dimensional spatial luminosity distribution for any simulated radio source in Section \ref{sec:SPATIALLY RESOLVED LOBE LOSSES}. In Section \ref{sec:FR-II SPATIAL LUMINOSITY EVOLUTION} we combine this resolved emissivity model with the dynamical model of \citet{Turner+2015} for lobed FR-II/I radio sources, {and investigate the validity of some simplifying assumptions}. A model for low-powered ``flaring jet'' FR-I sources is developed in Section \ref{sec:FLARING RADIO SOURCE MODEL}, and the potential for these ``flaring jet'' sources to explain the spectral features and compact nature of GPS and CSS sources is investigated. Finally, the effect of observing radio sources close to the detection limit in future radio continuum surveys is explored in Sections \ref{sec:UNDERSTANDING AGN FEEDBACK AND ENERGETICS}.

The $\Lambda \rm CDM$ concordance cosmology with $\Omega_{\rm M} = 0.3$, $\Omega_\Lambda = 0.7$ and $H_0 = 70 \rm\,km \,s^{-1} \,Mpc^{-1}$ \citep{Komatsu+2011} is assumed throughout the paper.

\section{LOBE EMISSIVITY MODEL}
\label{sec:LUMINOSITY MODEL}

We propose here a radio source luminosity model that can be applied to any dynamical model, be it analytical or numerical. This means that the radiative losses are not taken into account self-consistently in the evolution of the lobe pressure. The assumption is valid in a situation where the radiative losses do not change the lobe pressure significantly. This is always a good assumption if the lobe pressure is dominated by non-radiating particles such as protons. However, the assumption is also reasonable for purely leptonic plasmas; given electron injection indices steeper than $s = 2$, most of the energy is contributed by the low energy electrons. We confirm the validity of neglecting radiative losses a posteriori. Note that adiabatic losses are taken into account in numerical and published analytical dynamical models.

\citet{KA+1997} and \citet{KDA+1997} developed a model describing the temporal evolution of the powerful FR-II class of radio sources, including a full treatment of the radiative loss mechanisms for the lobe integrated luminosity.
However, these models make a number of simplifications to remain analytically tractable: (1) the external density profile is approximated by a single power law with a constant exponent \citep[inconsistent with X-ray cluster observations;][]{Vikhlinin+2006}, and (2) the radio lobes are assumed to expand in a self-similar manner and in the strong shock supersonic expansion limit.
In this section, we rederive their luminosity model for general radio sources which are not constrained by these restrictions. 

{We have eliminated many significant simplifying assumptions from previous radio source models; however we must still make some assumptions to keep the model analytically tractable including: (1) a spherically symmetric ambient medium, (2) a homogeneous magnetic field (excepting any gradients across the lobe) which contributes a constant and uniform fraction of the total internal pressure, and (3) no reacceleration of the synchrotron-emitting electrons by weak shocks or turbulence.
The \citet{Turner+2015} dynamical model can simulate radio AGN with each lobe expanding into different environments, and moreover the mathematical framework enables different density and temperature profiles to be assumed along every radial line from the central engine. However, without a compelling reason to choose a more complicated description in this work we adopt a spherically symmetric environment.
By contrast, we must defer to hydrodynamical simulations to understand inhomogeneities in the magnetic field \citep[but see e.g.][]{Hardcastle+2013} and equipartition factor (i.e. the ratio of energy densities in the field and particles). The assumption of a uniform and temporally constant fraction of total energy density in the magnetic field is verified by the three-dimensional MHD simulations of \citet[][their Figure 2]{Hardcastle+2014}. The turbulent nature of the flow is not expected to lead to a strong amplification of the magnetic field; it will however lead to fluctuations of the magnetic field and $\beta = 8\pi p /B^2$ of a factor of a few to ten \citep{Gaibler+2009}. This leads to a corresponding lack of noise in our model on all scales. 
We show throughout this work that despite these shortcomings our model still captures the main features of observed sources.}

\subsection{Unresolved, continuously injected electron model}

{The radio lobe luminosity is modelled throughout the source lifetime by determining the evolution of the spectral energy distribution of synchrotron-emitting electrons and of the lobe volume.}
{The spectral emissivity per unit volume $J(\nu)$ arising from a packet of synchrotron-emitting electrons expanding adiabatically with pressure evolution $p(t)$ is described in Appendix \ref{sec:appendix}.}
The luminosity of a lobe with constant emissivity is found using $L(\nu) = 4\pi J(\nu) V$ where $V$ is the emitting volume of the entire lobe. 
In general, the emissivity depends on the electron populations which are injected at different times in the radio source evolutionary history. The lobe has very different volumes at these times and we thus express the volume in integral form,

\begin{equation}
\begin{split}
V(t) &= \int_0^t \frac{dV(t_{\rm i})}{dt_{\rm i}} dt_{\rm i} = \int_0^t \frac{a_{\rm v}(t_{\rm i}) V(t_{\rm i})}{t_{\rm i}} dt_{\rm i} ,
\end{split}
\end{equation}

where the volume $V(t)$ of the entire lobe expands from some volume $V(t_{\rm i})$ when a given packet of electrons is injected {at time $t_{\rm i}$}. The lobe volume expands over this time as $V \propto t^{a_{\rm v}}$, where $a_{\rm v}$ can be a slowly varying function of time. The luminosity of the radio AGN lobe is then found from our emissivity and volume expressions:

\begin{equation}
\begin{split}
L(\nu, t) &= \int_0^t \frac{4\pi J(\nu, t, t_{\rm i}) a_{\rm v}(t_{\rm i}) V(t_{\rm i})}{t_{\rm i}} dt_{\rm i} .
\end{split}
\end{equation}

These expressions can be rewritten solely in terms of the key parameters in our model: lobe pressure $p$, volume $V$ and the equipartition factor $q$. That is,

\begin{equation}
\begin{split}
L(\nu, t) = K(s) &\nu^{(1 - s)/2} \frac{q^{(s + 1)/4}}{(q + 1)^{(s + 5)/4}} \\
&\times p(t)^{(s + 5)/4} V(t) \mathcal{Y}(t) ,
\end{split}
\label{luminosityloss}
\end{equation}

where the equation for the losses over the entire radio lobe at source age $t$ is given by

\begin{equation}
\begin{split}
\mathcal{Y}(t) = \int_0^t \frac{a_{\rm v}(t_{\rm i})}{t_{\rm i}} \left[\frac{p(t_{\rm i})}{p(t)} \right]^{1 - 4/(3\Gamma_{\rm c)}} \frac{V(t_{\rm i})}{V(t)} \left[\frac{\gamma_{\rm i}}{\gamma} \right]^{2 - s} dt_{\rm i} ,
\end{split}
\label{losseqn}
\end{equation}

{where $\gamma$ is the Lorentz factor at the present epoch and $\gamma_{\rm i}$ at the time of electron injection. 
In the case of impulsive electron injection at the present time, $t_{\rm i} = t$, the integral in this equation evaluates to unity, as expected in the case of no energy losses. Finally, the radio source specific constant $K(s)$ in Equation \ref{luminosityloss} is defined through}

\begin{equation}
\begin{split}
K(s) &= \frac{\kappa(s)}{{m_{\rm e}}^{(s + 3)/2} c (s + 1)} \left[\frac{e^2 \mu_0}{2(\Gamma_{\rm c} - 1)} \right]^{(s + 5)/4} \\
&\times \left[\frac{3}{\pi} \right]^{s/2} \left[\frac{\gamma_{\rm min}^{2 - s} - \gamma_{\rm max}^{2 - s}}{s - 2} - \frac{\gamma_{\rm min}^{1 - s} - \gamma_{\rm max}^{1 - s}}{s - 1}\right]^{-1} ,
\end{split}
\end{equation}

{where $c$ is the speed of light, $e$ and $m_{\rm e}$ are the electron charge and mass, $\mu_0$ is the vacuum permeability, $\Gamma_{\rm c}$ is the adiabatic index of the lobe, and $\kappa$ is a constant dependent solely on the injection index $s$ as defined in Equation \ref{kappas}. Additionally, $\gamma_{\rm min}$ and $\gamma_{\rm max}$ are the Lorentz factors corresponding to the low- and high-energy cut-offs to the injection electron energy distribution.}

\subsection{Modelling mock radio sources}
\label{sec:MODELLING MOCK RADIO SOURCES}

The radio lobe luminosity model derived in the previous section can be boot-strapped atop analytical or numerical lobe dynamical models for which the pressure and volume temporal evolution is known. 
We first illustrate this process with a radio lobe from the dynamical model of \citet{Turner+2015}. 
Their model fits a series of broken power-laws to the pressure profile, and evaluates lobe pressure, volume and axis ratio at each time-step. Here, we further fit a broken power-law time series to these model outputs, namely the volume and pressure of the lobe.
The temporal evolution of these simulation parameters is represented by the power-law exponents $a_{\rm p}$ and $a_{\rm v}$, where $p \propto t^{a_{\rm p}}$ and $V \propto t^{a_{\rm v}}$. These exponents are calculated by interpolating log-linearly at each time-step, $p(t_{\rm n+1}) = p(t_{\rm n}) (t_{\rm n+1}/t_{\rm n})^{a_{\rm p}(t_{\rm n})}$ and $V(t_{\rm n+1}) = V(t_{\rm n}) (t_{\rm n+1}/t_{\rm n})^{a_{\rm v}(t_{\rm n})}$.
Here, the time-steps run backwards in time from the source age to the time of electron injection as for the Lorentz factor recurrence relation in Equation \ref{injectionlorentz}, i.e. $t_0 = t$ and $t_{\rm N} = t_{\rm i}$. The Lorentz factor at the time of observation and the time of electron injection can thus be calculated.
The precision of the loss mechanism integral is increased by interpolating the data points from the dynamical model enabling higher cadence sampling in injection time.

\subsection{Evolutionary tracks}
\label{sec:EVOLUTIONARY TRACKS}

Radiative loss mechanisms, including adiabatic, synchrotron and inverse-Compton, can change the jet power--luminosity relation by over an order of magnitude as a radio source ages \citep{KDA+1997, Blundell+1999, Manolakou+2002, Shabala+2013, Turner+2015}.
\citet{KDA+1997} proposed a method for simulating the losses in self-similar FR-II radio lobes expanding into environments modelled by a single power-law density profile. 
In our prescription both the density and temperature profiles can vary with radius in an arbitrary manner, and differ for each radio lobe \citep{Turner+2015}. This enables observed environments of a real source or any arbitrary environments to be used as the basis for the radio source dynamics.

\begin{figure}
\begin{center}
\includegraphics[width=0.48\textwidth]{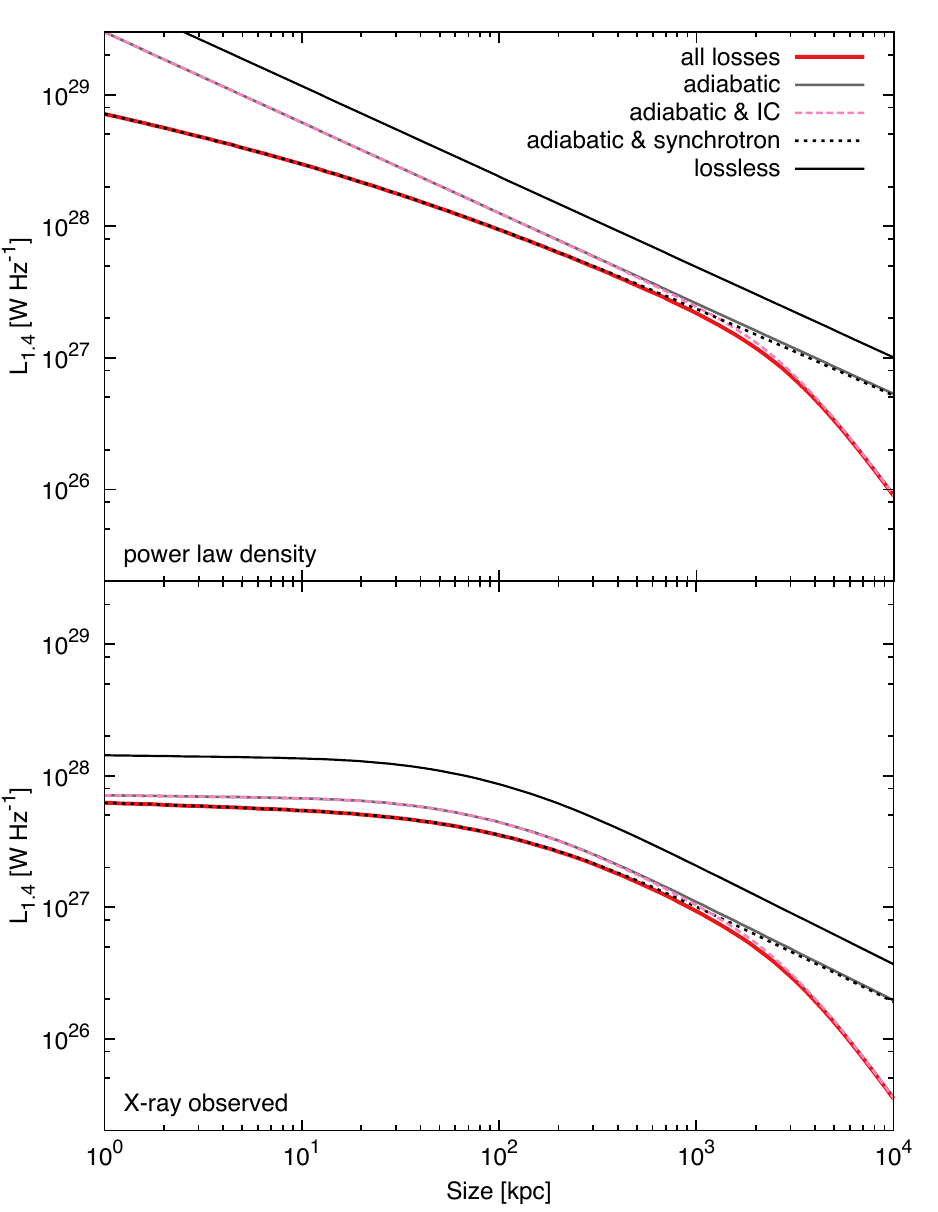} 
\end{center}
\caption{Top panel -- Luminosity-size tracks for a $1.3 \times 10^{39} \rm\, W$ jet power radio source at $z = 0.5$, expanding into an environment with density $\rho = 7.2\times 10^{-22} \rm\, kg\,m^{-3}$ at a radius of $2 \rm\, kpc$ falling off with a power-law profile with constant exponent $\beta = 1.9$. Five profiles are plotted, showing the luminosity for the lossless case, with only adiabatic losses, with adiabatic and inverse-Compton losses, with adiabatic and synchrotron losses, and with all loss mechanisms. Bottom panel -- Same as top panel but for an environment based on X-ray cluster observations with the same sub-halo mass and virial radius as simulated for the power-law density profile.}
\label{fig:LDtrackbeta}
\end{figure}

The differences in the evolutionary tracks modelled using either the power law \citet{KDA+1997} environments and a more complex environment allowed by our model 
are shown in Figure \ref{fig:LDtrackbeta}. The single-power law assumption leads to a perfect power-law fall-off in luminosity for the lossless case and when including losses due to the adiabatic expansion of the self-similar lobe. Losses due to synchrotron radiation preferentially reduce the luminosity at early times whilst losses due to the inverse-Compton scattering of cosmic microwave background radiation are only noticeable after $\sim 100\rm\,Myrs$. By contrast, fitting a declining density profile to the host sub-halo environment yields significant curvature in the lossless luminosity with values two-orders of magnitude lower in the cluster core regions. The adiabatic losses no longer yield a constant difference in luminosity either since we are not assuming self-similar expansion.


\section{SPATIALLY RESOLVED LOBE LOSSES}
\label{sec:SPATIALLY RESOLVED LOBE LOSSES}

The method described thus far for calculating the lobe luminosity implicitly assumes synchrotron emitting electrons are uniformly injected at locations throughout the entire lobe volume. An alternative interpretation is that the model assumes the radio source is unresolved and these injection locations cannot be determined. Either way, when considering the emission from a resolved source, in which all\footnote{Shocks throughout the radio lobe may re-accelerate electrons back up to higher energies, however here we assume this process to have only a minor effect on the electron energy distribution. This is likely to be true particularly for the subsonic flow of plasma beyond the flare point in FR-Is .} electrons are accelerated at a compact hotspot in FR-IIs or flare point in FR-Is, a different approach must be taken.

\subsection{Spatially distributed and impulsively injected electron model}

The different injection locations and pressure distributions seen in type I and II radio sources lead to different evolution of their electron populations.
The volume and pressure evolution used for the electron populations in lobed FR-II radio sources continues to be described by the large-scale dynamics of the lobe. This assumption is valid because the high sound speed in the lobe brings the entire lobe into approximate pressure equilibrium \citep[corrections can be applied for the higher pressure expected at the hotspot;][]{Kaiser+1999}. The volume initially occupied by each injection of electrons should similarly increase in proportion to volume of the entire lobe. The situation is different for the smoke-like plumes of plasma seen in ``flaring jet'' FR-Is, which are in approximate pressure equilibrium with their surroundings and slowly coast down the declining pressure gradient beyond the flaring point. The pressure and volume occupied by an electron packet at this acceleration site remains constant throughout the source lifetime. The subsequent evolution can be characterised using an FR-I dynamical model \citep[see Section \ref{sec:FLARING RADIO SOURCE MODEL};][]{Luo+2010}.

In the following, we assume that the losses at a given location in the lobe result from electrons all accelerated at the same time.
These synchrotron-emitting electrons traverse the distance from the site of particle acceleration to their present time location as part of the large scale motions of plasma in the lobe (e.g. backflow from the hotspot for FR-II sources, or forward flow from the flare point for twin-jet FR-I sources). 
The emission at individual locations in the lobe now arises solely from a single injection time. The volume occupied at the present time by an electron packet injected between $t_{\rm i}$ and $t_{\rm i} + dt_{\rm i}$ is given by

\begin{equation}
\begin{split}
dV(t, \textbf{r}) &= \int_0^t \delta(t_{\rm i}' - t_{\rm i}(\textbf{r})) \frac{dV(t_{\rm i}')}{dt_{\rm i}'} dt_{\rm i}' ,
\end{split}
\label{resolvedvolume}
\end{equation}

where $\textbf{r}$ is the radial position vector of the present-time electron emission region from the acceleration site. For lobed FR-II type radio sources the differential volume element of the plasma comprising electrons of the age at position $\textbf{r}$ can be written more neatly as

\begin{equation}
\begin{split}
dV(t, \textbf{r}) &= \frac{a_{\rm v}(t_{\rm i}(\textbf{r})) V(t_{\rm i}(\textbf{r}))}{t_{\rm i}(\textbf{r})} dt_{\rm i} .
\end{split}
\label{resolvedlosses}
\end{equation}

This form enables the integrated luminosity to be easily calculated using numerical integration.

By contrast, in low-powered flaring jet FR-I sources the flare point acceleration site is stationary and unchanging in pressure throughout the source lifetime (we further discuss flaring jet dynamics in Section \ref{sec:FLARING RADIO SOURCE MODEL}). The volume of the injected electron packets at the time of acceleration is thus constant; i.e. $V(t_{\rm i}(\textbf{r})) = V_{\rm i}$ and hence $dV_{\rm i}/dt_{\rm i} = 0$, where the volume $V_{\rm i}$ is related to the size of the flare point. The synchrotron-emitting electrons propagate outwards from the flaring point towards the maximal extent of the radio plume, reaching their present-time location $\textbf{r}$ in time $t_{\rm e}(\textbf{r}) = t - t_{\rm i}(\textbf{r})$. The volume integral in Equation \ref{resolvedvolume} is therefore recast in terms of this electron travel time $t_{\rm e}$ for low-powered sources.

The luminosity of the lobe section at radial position $\textbf{r}$ is now given by

\begin{equation}
\begin{split}
dL(\nu, t, \textbf{r}) &= 4\pi J(\nu, t, t_{\rm i}(\textbf{r}))\:\! dV(t, \textbf{r}) .
\end{split}
\end{equation}

Then as for the unresolved radio lobe, upon substituting in the expressions for the lobe volume and emissivity we obtain

\begin{equation}
\begin{split}
dL(\nu&, t, \textbf{r}) = K(s) \nu^{(1 - s)/2} \frac{q^{(s + 1)/4}}{(q + 1)^{(s + 5)/4}} dV(t, \textbf{r}) \\
&\times p(t, \textbf{r})^{(s + 5)/4} \left[\frac{p(t_{\rm i}(\textbf{r}))}{p(t, \textbf{r})} \right]^{1 - 4/(3\Gamma_{\rm c})} \left[\frac{\gamma_{\rm i}(\textbf{r})}{\gamma} \right]^{2 - s}  ,
\end{split}
\label{resolvedluminosity}
\end{equation}

where in lobed FR-IIs, $p(t, \textbf{r})$ and $p(t_{\rm i}(\textbf{r}))$ are the modelled lobe pressure at the present time and the time of electron injection respectively. 
Low-powered flaring jet radio sources, which do not have constant pressure lobes, instead take $p(t_{\rm i}(\textbf{r}))$ as the flare point pressure and $p(t, \textbf{r})$ as the pressure of an electron packet which has propagated for time $t_{\rm e}(\textbf{r})$ along the plume. This equation can alternatively be applied to the local variations in the pressure of powerful FR-IIs predicted by hydrodynamical simulations.

\subsection{Temporal evolution of synchrotron-emitting electrons}

The entire lobe radio luminosity is reduced at higher frequencies due to the power-law fall-off in the energy of the electron population, i.e. $L_{\nu} \propto \nu^{(1 - s)/2}$ for $s \sim 2$-3. The frequency-dependent nature of the loss mechanisms implicitly included in Equation \ref{resolvedluminosity} further reduces the intensity of emission at higher frequencies as the synchrotron-emitting electrons age. This can be demonstrated by combining Equations \ref{ngamma} and \ref{injectionlorentz} for the number of synchrotron-emitting electrons injected at some time $t_{\rm i}$ which are emitting at frequency $\nu \propto \gamma^2$ at the time of observation $t$. That is,

\begin{equation}
N(\nu, t, t_{\rm i})\, d\nu = N_1 (t^{-a_{\rm v}/3} - a_2(t, t_{\rm i}) k_1 \nu^{1/2})^{s - 2} \,d\nu ,
\label{mininject}
\end{equation}

\begin{figure}
\begin{center}
\includegraphics[width=0.48\textwidth]{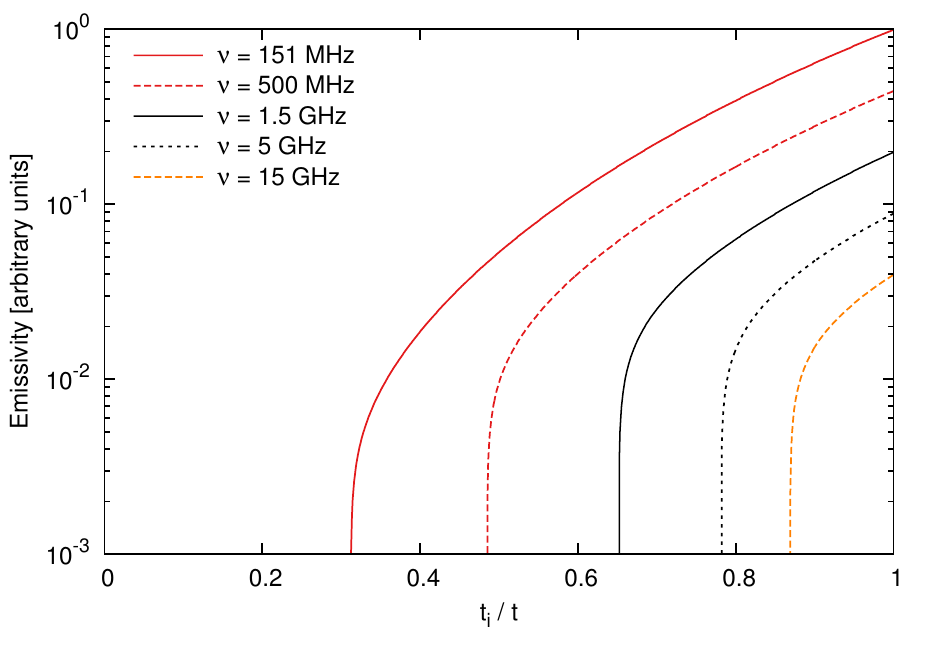}
\end{center}
\caption{Synchrotron-emitting electrons emissivity as a function of the electron injection time (for a typical $100\rm\, Myr$ source at $z = 0.5$; refer to Figure \ref{fig:LDtrackbeta}) for five typical observing frequencies of $151\rm\, MHz$, $500\rm\, MHz$, $1.5\rm\, GHz$, $5\rm\, GHz$ and $15\rm\, GHz$. Electrons injected at younger than approximately half the source age for this example source emit no radiation at the time of observation.}
\label{fig:tinj}
\end{figure}

where $N_1$ is some arbitrary constant of proportionality and $k_1$ is defined such that $\gamma = k_1 \nu^2$ (see Equation \ref{gamma}). The function $a_2(t, t_{\rm i})$ derived by \citet{KDA+1997} is not a function of frequency, and increases as $t_{\rm i}/t$ gets very small. The number density of electrons emitting synchrotron radiation at the present time at a given frequency may therefore reduce to zero for suitably small $t_{\rm i}/t$, i.e. $a_2(t, t_{\rm i}) \geqslant t^{-a_{\rm v}/3}/(k_1\nu^{1/2})$. That is, there is a cut-off injection age below which synchrotron-emitting electrons cannot emit at a given rest frequency at the present time. Moreover, this cut-off injection age is expected to be a strong increasing function of frequency, i.e. $t_{\rm i\, cut} = f(\nu^{1/2})$. 
The loss function for a typical radio source is shown in Figure \ref{fig:tinj} for five standard observing frequencies. 
The losses due to the inverse-Compton upscattering of CMB photons are a strong function of both the age of synchrotron-emitting electrons and the source redshift. The cut-off injection ages shown are therefore likely to vary greatly for different sources.

\subsection{Spatial distribution of electron ages}

The synchrotron age distribution throughout radio AGN lobes is likely a complicated function of position, however spectral ageing does identify a general linear increase in age away from the hotspots of FR-II class sources \citep[e.g.][]{Jamrozy+2008, Harwood+2013}. This increase in synchrotron age arises due to a backflow of plasma carrying freshly injected electrons away from the hotspot towards the AGN core in FR-IIs. In unconfined FR-Is, a similar role is played by forward flow from outflowing jets. These observations, although informative, can only estimate the average electron age in a three-dimensional slice along the line-of-sight. Hydrodynamical simulations or some other physically based argument are required to properly characterise the spatial distribution of synchrotron-emitting electrons. In this work, we assume the synchrotron age increases with radial distance from the injection site at some speed $b(\textbf{r}) + db(\textbf{r})$, guided by hydrodynamical simulations for FR-IIs and analytic theory for FR-Is. Typical values for this flow rate estimated using spectral ageing lie somewhere in the range $b = 0.01$-$0.1 \rm\, pc/yr$ \citep{Jamrozy+2008}.

\section{POWERFUL RADIO GALAXIES}
\label{sec:FR-II SPATIAL LUMINOSITY EVOLUTION}

The spatial distribution of energy losses in the synchrotron-emitting electron population is typically not considered in analytical models \citep[e.g.][]{KDA+1997} and hydrodynamic simulations (e.g. \citealt{HK+2013, Hardcastle+2014}; but see e.g. \citealt{Jones+1999}). This, in turn, leads to oversimplified predictions for the spatial distribution of radio continuum emissivity.
The framework derived in Section \ref{sec:SPATIALLY RESOLVED LOBE LOSSES} is employed in this section using results from the \citet{Turner+2015} model.
The detailed spatial distribution of the synchrotron-emitting electrons in FR-II type sources after they are injected into the lobe is modelled using hydrodynamical simulations.

\subsection{Hydrodynamical simulations}

The hydrodynamical simulations in this work use the freely available PLUTO\footnote{\url{http://plutocode.ph.unito.it/}} code, version 4.2, described by \citet{Mignone+2007}. We model the radio lobes in two-dimensional cylindrical coordinates to ensure an ample number of cells along the length of the jet, and in particular at dynamically complex regions such as the working surface. Modelling of the lobe using cylindrical coordinates in the PLUTO code ensures the correct three-dimensional spatial dependency of physical parameters as the fluid from the jet terminal shock disperses along the lobe.
\citet{Krause+2001} find a resolution of greater than approximately 50 pixels across the jet width yields internal energies, axial momentum, mass and average bow shock velocities within $5\%$ of the convergent true value. This study was performed in two-dimensions with the PLUTO code. The simulations in our work will therefore use either a low-resolution grid of $750\times1800$ with 60 pixels across the jet for our sensitivity analyses, or a high-resolution map with $1500\times3600$ and 120 pixels across the jet for the final analysis. The jets of powerful FR-II radio sources are assumed to quickly collimate; we therefore inject a constant width jet into the simulation grid to reduce computation time.

\subsection{Tracing the rate of backflow}
\label{sec:Tracing the rate of backflow}

\begin{table*}
\begin{center}
\caption[]{Hydrodynamical simulations run in this work using the PLUTO code. The reliability of the results from the main simulation {\sc Base} are tested for changes to the Mach number (column 2), jet-environment density ratio (column 3), the profile assumed for the host gas density (column 5), and the resolution (column 6). The ages to which the radio lobe is simulated are also included (column 4).}
\label{tab:plutosims}
\setlength{\tabcolsep}{10pt}
\begin{tabular}{cccccc}
\hline\hline \vspace{-0.35cm}\\
Simulation&$\mathcal{M}$&$A_\rho$&Age&Profile&Resolution\vspace{-0.035cm}
\\
&&&\!\!\!(sim.\,unit)\!\!\!&&(cells, $r_{\rm cyl} \times z$)\vspace{0.035cm}
\\
\hline \vspace{-0.35cm}\\
{\sc Base}	&	100	&	0.01	&	8,\,9,\,10	&flat&	$750\times1800$\\
{\sc Jet\:Power}	&	50	&	0.01	&	8,\,9,\,10	&flat&	$750\times1800$\\
{\sc Axis\:Ratio}	&	100	&	0.001	&	8,\,9,\,10	&flat&	$750\times1800$ \vspace{0.035cm}\\
\hline \vspace{-0.35cm}\\
{\sc Density}	&	100	&	0.01	&	10	&$\rho = 1/r$&	$750\times1800$\\
{\sc Resolution}	&	100	&	0.01	&	10	&flat&	$1500\times3600$ \vspace{0.035cm}
\\
\hline
\end{tabular}
\end{center}
\end{table*}

The spatial distribution of synchrotron-emitting electrons throughout the radio lobe is simulated by injecting tracer fields into the jet (and thus lobe) at regular intervals and tracking their motion. The location of these tracer fields within the lobe at any given radio source age is used to calculate the contribution to the emission in any region from the various age electrons. Tracer fields are injected every 0.2 simulation units up to the source age of $t=10$ simulation units. The use of the hydrodynamical simulations is maximised by examining the distribution of the tracer field at source ages of $t=8$, 9 and 10 simulation units. Hydrodynamical simulations are run for three high-powered radio sources using the low-resolution grid, with the key parameters varied between runs. The external Mach number of the jet (jet velocity\,/\,ambient sound speed) $\mathcal{M}$, which is directly related to the jet power, is simulated for a base case of $\mathcal{M} = 100$ and varied to a lower $\mathcal{M} = 50$. The ratio of the density in the jet and ambient medium, related to the lobe axis ratio, is simulated for $A_\rho = 0.01$ and 0.001, as listed in Table \ref{tab:plutosims}. The single high-resolution grid is used to confirm the stability of our results. Note that the non-dimensionalised parameters in these simulations can be scaled (one-to-many) into equivalent physical properties following the method of \citet{Krause+2012}. A further simulation was run with a non-constant slope density profile to confirm the backflow of plasma within the high sound speed radio lobe is unaffected by environment as theoretically expected.

\begin{figure}
\begin{center}
\includegraphics[width=0.495\textwidth]{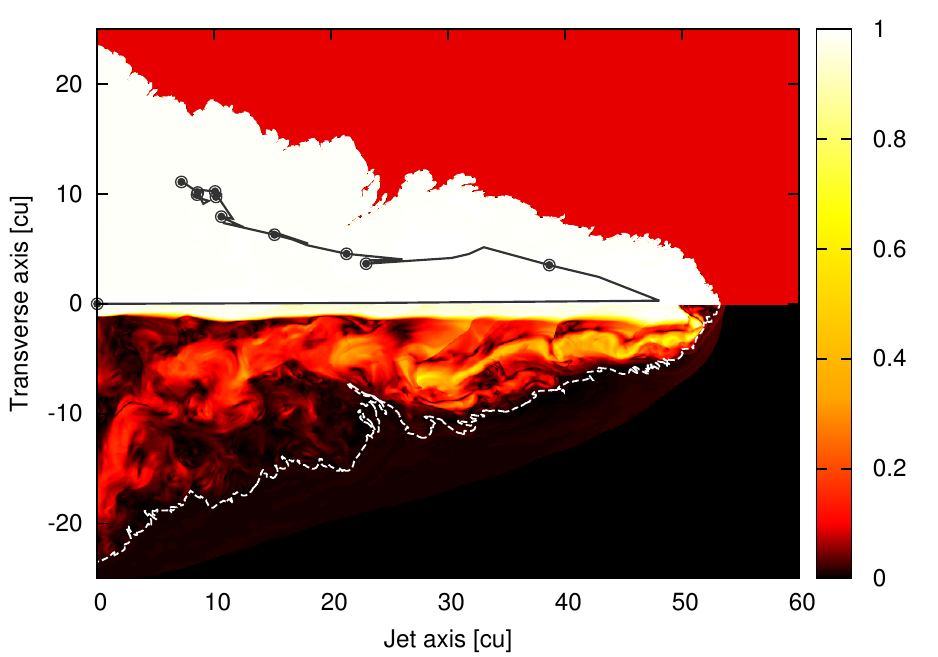}
\end{center}
\caption{Hydrodynamical simulation ({\sc Base} run) of backflow of packets of synchrotron emitting electrons from the hotspot injection site towards the core. The bottom half of the plot shows the magnitude of the velocity of the lobe plasma as it travels both along the jet with velocity scaled to unity, then back towards the core once injected into the lobe. The white dashed line marks the contact discontinuity between the lobe plasma and the shocked ISM. The solid black line in the top half of the plot shows the average location of the tracer fields injected at the current source age (point at start of jet) through to the packet injected at the instantiation of the source. The points are spaced along the average backflow path every one-tenth of the source age of $t=10$ simulation units.}
\label{fig:hydrosim}
\end{figure}

The backflow of lobe plasma simulated using PLUTO for our {\sc Base} case is plotted in Figure \ref{fig:hydrosim}. The fluid injected at the jet terminal hotspot initially flows smoothly back towards the core, but the flow becomes more turbulent after travelling back less than half of the lobe length. The tracers become somewhat dispersed thereafter, though their progression along the length of the lobe is still apparent. The average location of the tracer field is plotted in the top half of the figure. This shows an initially rapid progression towards the core before the tracer accumulates at approximately one-fifth the lobe length (from the core). The tracer of course spreads out well beyond this mean position; the distribution of the tracer out to the $2\sigma$ noise level is therefore taken when modelling the spatial distribution of the synchrotron-emitting electrons. 

\begin{figure}
\begin{center}
\includegraphics[width=0.45\textwidth]{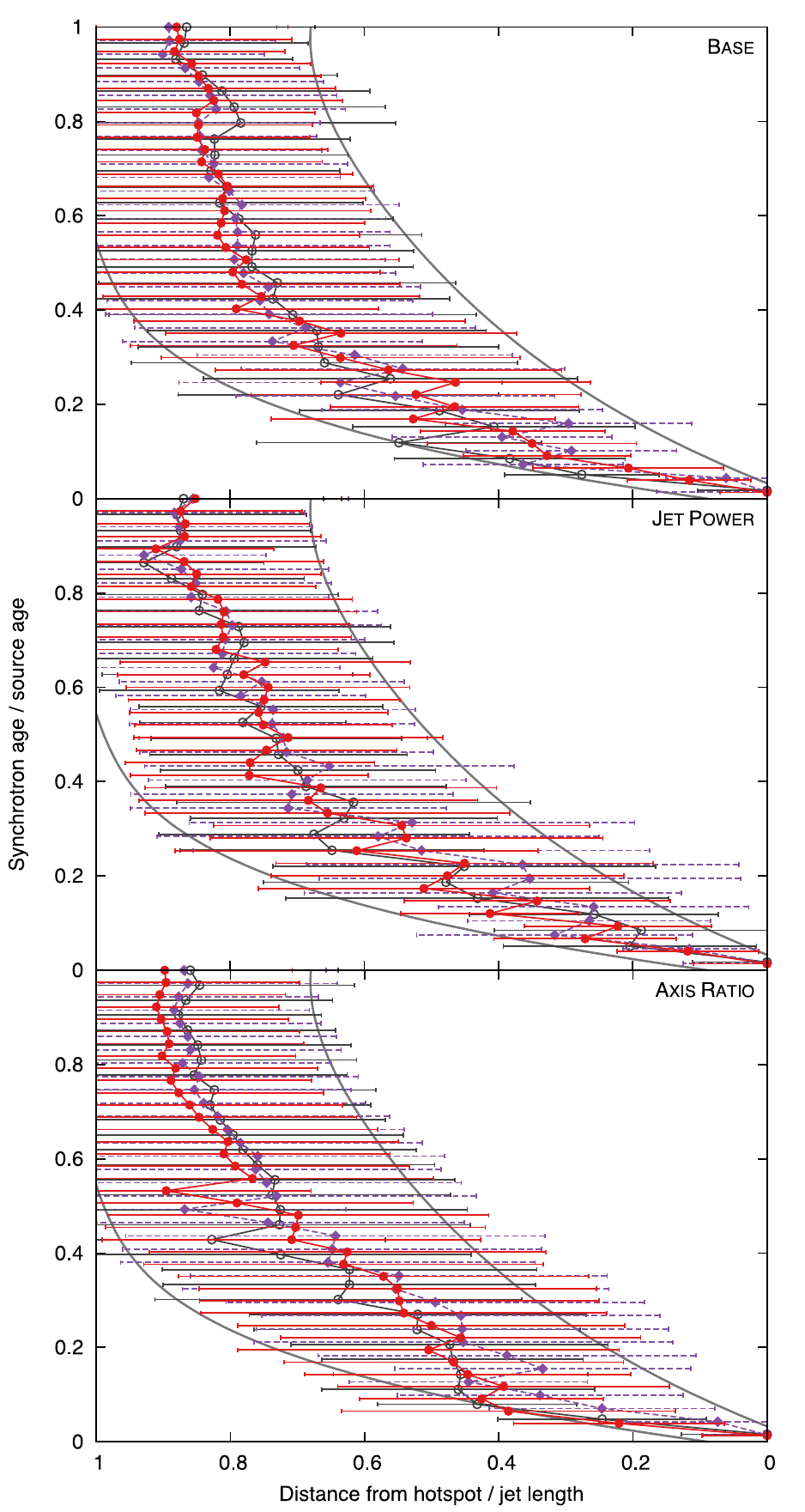}
\end{center}
\caption{Age of synchrotron-emitting electrons as a function of distance from the hotspot injection site for the three low-resolution simulations (as indicated in panels). The points show the mean position of the tracer field for a packet of injected electrons at times $t=8$ in grey (open circles), $t=9$ in purple (triangles), and $t=10$ simulation units in red (filled circles). The error bars show the $1\sigma$ spread in the tracer field along the radial vector $\textbf{r}$. Polynomial fits to the $1\sigma$ spread in the tracer field are plotted with solid grey lines, showing that the bulk of freshly injected electrons emit close to the hotspot.}
\label{fig:injectprofile}
\end{figure}

The injection age for the synchrotron-emitting electrons is taken as the time they begin to spread radially, which we confirmed visually coincides with passing through the terminal shock of the jet. The spread of synchrotron-emitting electrons of each injection age is calculated along a radial axis extending outwards from the hotspot injection site, i.e. we are only concerned with travel time from the hotspot. The spatial distribution of the various age electrons along the axis is plotted in Figure \ref{fig:injectprofile} for each of the three low-resolution simulations at $t=8$, 9 and 10 simulation units. The radial axis is scaled to the total lobe length (i.e. the length of the lobe is one in simulation units) and the age of the electrons is scaled to the source age (i.e. freshly injected electrons have zero age, the oldest electrons have age unity). The fits to the mean location of the tracer are consistent for each simulation (solid red points) suggesting these results are scalable to any high-powered FR-II radio source. Similarly the plotted $1\sigma$ spread in the tracer is comparable for each simulation. Importantly, this figure shows that the bulk of freshly injected electrons emit close to the hotspot, and thus much of the lobe may not be visible in sources where the loss mechanisms prevail.

The average location of the electron packets as a function of synchrotron age is determined from these simulations by fitting a fourth order polynomial to the data. The spread is similarly characterised by fitting a fourth order polynomial to the distribution of the standard deviations. These two fits are:

\begin{subequations}
\begin{eqnarray}
\tilde{\textbf{r}}(\tau) &=& 3.5\tau -6.6{\tau}^2 + 6.4{\tau}^3 - 2.4{\tau}^4 \\
d\tilde{\textbf{r}}(\tau) &=& 0.1 + 1.6\tau -4.2{\tau}^2 + 4.3{\tau}^3 - 1.6{\tau}^4 ,
\end{eqnarray}
\label{electrons}
\end{subequations}

where the non-dimensionalised synchrotron age and radial vector are $\tau = 1 - t_{\rm i}/t$ and $\tilde{\textbf{r}}$ respectively. For a typical source age of $10\rm\, Myrs$ and distance from the hotspot injection site of $10\rm\, kpc$ this yields a backflow velocity of approximately $0.02\rm c$ \citep[c.f.][]{Alexander+1987} until the electrons begin the accumulate near the core. This relation is applied in our spatially resolved loss model of Section \ref{sec:SPATIALLY RESOLVED LOBE LOSSES} to more realistically estimate the multi-frequency emissivity of spatially resolved radio sources.

\subsection{Three-dimensional emissivity distribution}

\begin{figure}
\begin{center}
\includegraphics[width=0.45\textwidth]{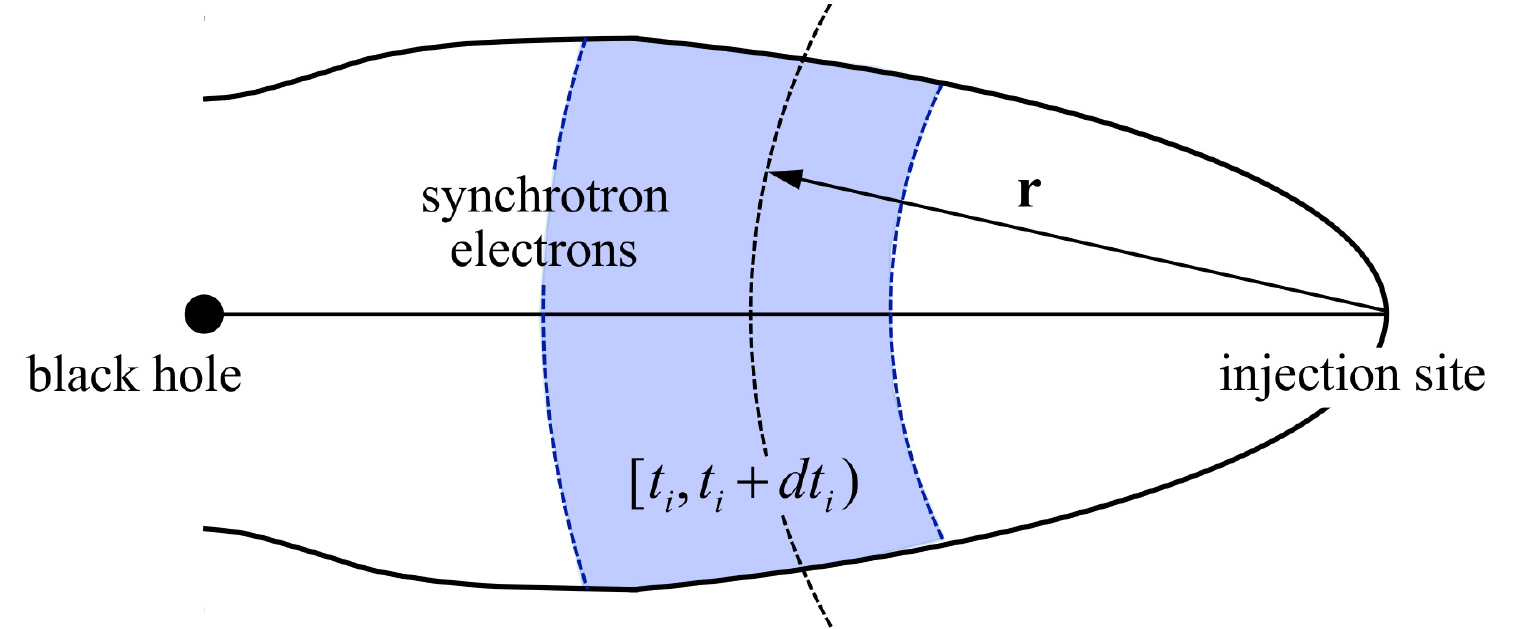}
\end{center}
\caption{Schematic of method used to apply spatial loss model to the output of the \citet{Turner+2015} dynamical model. At the time-step of interest the spatial distribution of the synchrotron-emitting electrons is calculated, e.g. the blue shading shows the spread of electrons injected at times $[t_{\rm i}, t_{\rm i} + dt_{\rm i})$. The spread of these electrons and their emissivity are both calculated in terms of the radial position vector $\textbf{r}$.}
\label{fig:diagram}
\end{figure}

The prescription for calculating the three-dimensional emissivity distribution of arbitrarily shaped radio lobes was detailed in Section \ref{sec:SPATIALLY RESOLVED LOBE LOSSES}. The shape of the lobe occupied by synchrotron emitting plasma is modelled using the outputs from the \citet{Turner+2015} dynamical model. In this work, we only use this model to represent powerful FR-II sources whose jets extend to a terminal hotspot at the end of the radio lobe\footnote{This model can produce radio sources with jets that are disrupted before reaching the end of the lobe, however we do not explore the effect of these different electron injection locations at this time.}. The origin of the radial vector $\textbf{r}$ is therefore positioned along the jet axis at the maximal extent of each lobe in our simulated radio sources, as shown in Figure \ref{fig:diagram}. The spatial loss model requires knowledge of only the evolutionary history of the spatial pressure distribution and the volume of the lobe as a whole. The pressure of high-powered FR-IIs is approximately constant across the lobe at any given time due to the high internal sound speed, and we can therefore assume an average lobe value. The final location of the synchrotron-emitting electrons is therefore unimportant to the calculation of the emissivity from a given age electron packet, only the effect they have on the emissivity distribution of the resolved lobe. The contribution to the total lobe emissivity from synchrotron-emitting electrons in each injection age range $[t_{\rm i}, t_{\rm i} + dt_{\rm i})$ can therefore be calculated using Equations \ref{resolvedlosses} and \ref{resolvedluminosity}, based on the lobe pressure and volume taken from \citet{Turner+2015}.

The location of a given age packet of electrons within an FR-II radio source is determined from: (1) their location within the non-dimensionalised lobe modelled using hydrodynamical simulations, and (2) scaled to the physical size and age of the radio lobe using the \citet{Turner+2015} model. The radio source emissivity is visualised by constructing a three-dimensional grid of pixels. Any pixels determined to lie outside the lobe defined at any given time-step by the \citet{Turner+2015} model are flagged and excluded from the emissivity calculation. The synchrotron-emitting electrons of each age are distributed through the lobe grid following a Gaussian profile described by the mean radius and standard deviation fitted by Equation \ref{electrons}. This Gaussian distribution is truncated at the $2\sigma$ level. 
The synchrotron emissivity arising from a given pixel is then calculated by summing the contribution from every age electron packet allocated to that cell.


\subsection{Two-dimensional emissivity}
\label{sec:Two-dimensional emissivity}

The observed surface brightness distribution from real sources is the integral of the luminosity from all volume elements along a given line-of-sight. The two-dimensional brightness distribution is calculated by simply summing the contribution from every cell along the depth of the source, assuming the low density lobe plasma is optically thin \citep{Peterson+1997}. In this work, for simplicity we assume that the radio lobes are viewed perpendicular to the axis of the jets. 
The surface brightness distribution can easily be determined for other viewing angles by rotating the line-of-sight through the three-dimensional grid, however the Doppler boosted jet emission will begin to dominate as the jet axis approaches the line-of-sight.

\begin{figure}
\begin{center}
\includegraphics[width=0.485\textwidth]{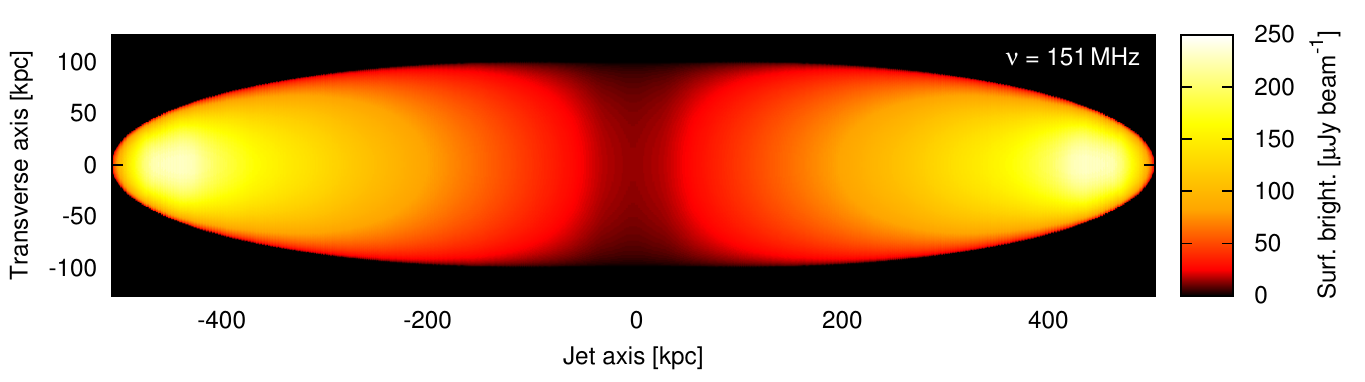}\\
\includegraphics[width=0.485\textwidth]{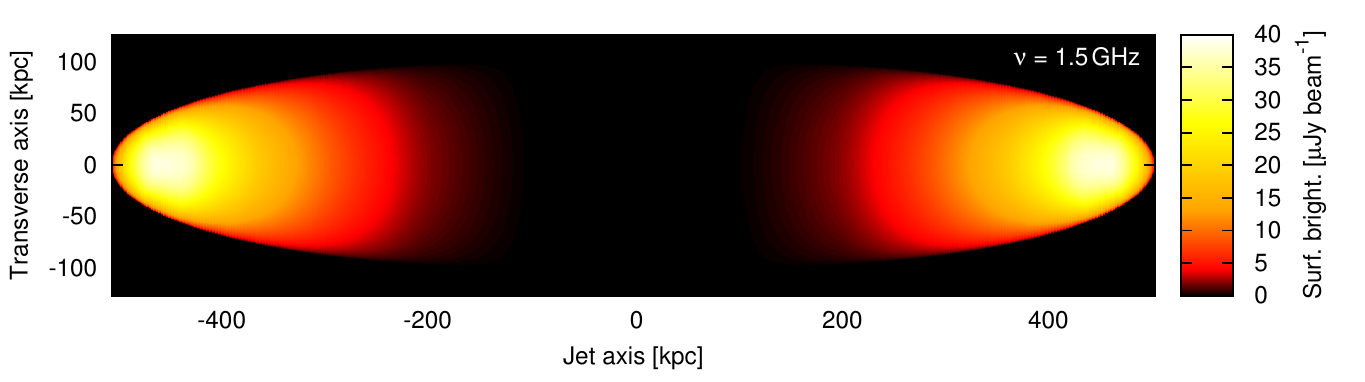}\\
\includegraphics[width=0.485\textwidth]{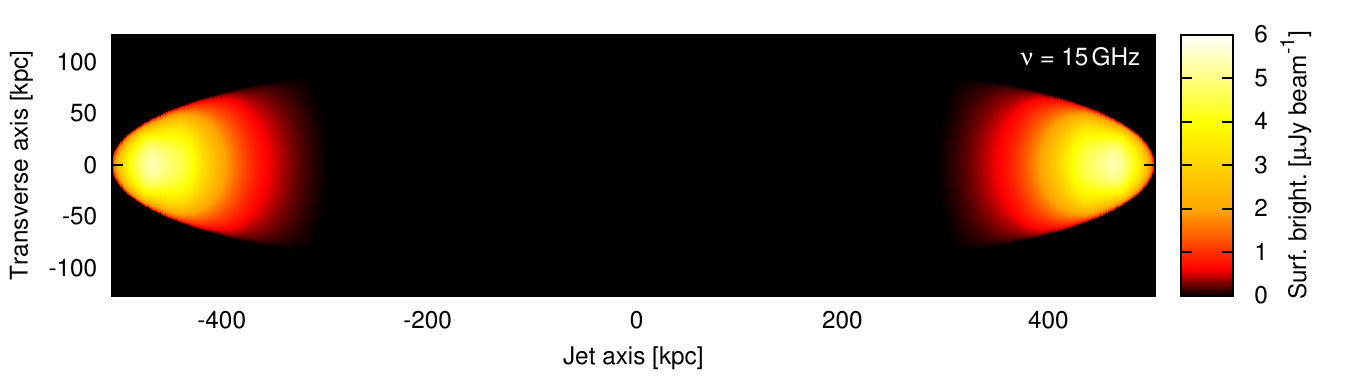}
\end{center}
\caption{Radio source emissivity as a function of lobe position including all loss mechanisms for three typical observing frequencies. The surface brightness is calculated assuming a $1\rm\, arcsec^2 (\lambda/\lambda_{1.5\rm\, GHz})^2$ beam. The top plot shows $151\rm\,MHz$ radio emission, the middle plot is for $1.5\rm\,GHz$, and the bottom plot shows $15\rm\,GHz$ emission. The modelled source has a jet power of $Q = 10^{40} \rm\, W$, age of $t = 100\rm\, Myrs$, inhabits a $10^{14} \rm\, M_\odot$ halo mass environment at $z = 1$, expands with an initial axis ratio of $A = 4$ and has an electron energy injection index of $s = 2.4$.}
\label{fig:LDspatial}
\end{figure}

The emissivity distribution modelled for a typical FR-II is shown in Figure \ref{fig:LDspatial} at three standard observing frequencies. The integrated lobe radio luminosity is lower at higher frequencies due to the power-law fall-off in the number counts of high-energy electrons. The frequency dependent nature of the loss mechanisms also reduces the timescales over which electrons emit detectable radiation at a given frequency. The size of the emitting region of the radio lobe therefore decreases with increased observing frequency, in addition to the overall reduction in lobe luminosity. Moreover, for an ellipsoidal radio lobe assumed by our model, the observed axis ratio will appear to be greater at higher frequencies due to the lack of emission at the wider, older parts of the lobe near the nucleus. {The size of the emission region reduces in this manner with both increasing redshift and source age due primarily to the inverse-Compton loss mechanism.}

\subsection{Spatial radio SED analysis}
\label{sec:SPATIAL RADIO SED ANALYSIS}

The resolved emissivity model (Section \ref{sec:SPATIALLY RESOLVED LOBE LOSSES}) can be used to calculate the emissivity in the lobe as a function of frequency and position; the spectral energy distribution (SED) can thus be derived at any location in the lobe. The two-dimensional distribution of spectral ages observed in resolved radio sources can be simulated by fitting an appropriate spectral model to the modelled radio SED in small cells across the mock radio lobe.
Spectral age measurements \citep[e.g.][]{Jamrozy+2008, Harwood+2013} are typically made using an impulsive injection model. The electrons in each cell (or slice) of the lobe are assumed to be injected at approximately the same time, though in reality likely cover a range in electron ages (see Figure \ref{fig:injectprofile}). Regardless, in this analysis an impulsive injection model will be used to enable direct comparison with existing observations.
The radio SED must therefore be fitted using either the Jaffe-Perola \citep[JP;][]{Jaffe+1973} or Kardashev-Pacholczyk \citep[KP;][]{Kardashev+1962} models. These models both assume that the entire electron population is accelerated at time $t = 0$ then undergoes radiative losses thereafter. The validity of using these impulsive injection models on the well-mixed lobes of FR-IIs can be tested by comparing their spectral age estimates to the source dynamical age.

The lobed FR-II/I dynamical model of \citet{Turner+2015} and the new resolved emissivity model are tested using the {spatial distribution of spectral ages} measured for real sources. 
In particular, we reproduce the spectral age distribution of 3C436 observed by \citet{Harwood+2013}. The intrinsic properties of 3C436 are fitted using the Bayesian algorithm of \citet{Turner+2016} based upon multi-frequency luminosity observations \citep{Laing+1980}  with the lobe size and axis ratio taken from the $8.4\rm\, GHz$ FITS file \citep{Hardcastle+1997}. The host environment of 3C436 is modelled here using semi-analytic galaxy evolution \citep[SAGE][]{Croton+2006} models informed by the host stellar mass ($M_\star = 10^{11.8} \rm\, M_\odot$). We obtain a jet power of $10^{39}\rm\, W$, source age of $37\rm\, Myrs$ and an equipartition factor of $B/B_{\rm eq} = 0.28$. 
{The observed spectral ages are recalculated assuming the dynamical model magnetic field strength to enable a more direct comparison between the electron flow predicted by our model and that of real sources. The spectral age is directly proportional to the magnetic field strength and break frequency; differences between the  spectral age distributions must therefore be due to: (1) inhomogeneities in the magnetic field, (2) the reacceleration of synchrotron-emitting electrons, or (3) discrepancies in the electron flow model.}

\begin{figure*}
\begin{center}
\includegraphics[width=0.36\textwidth]{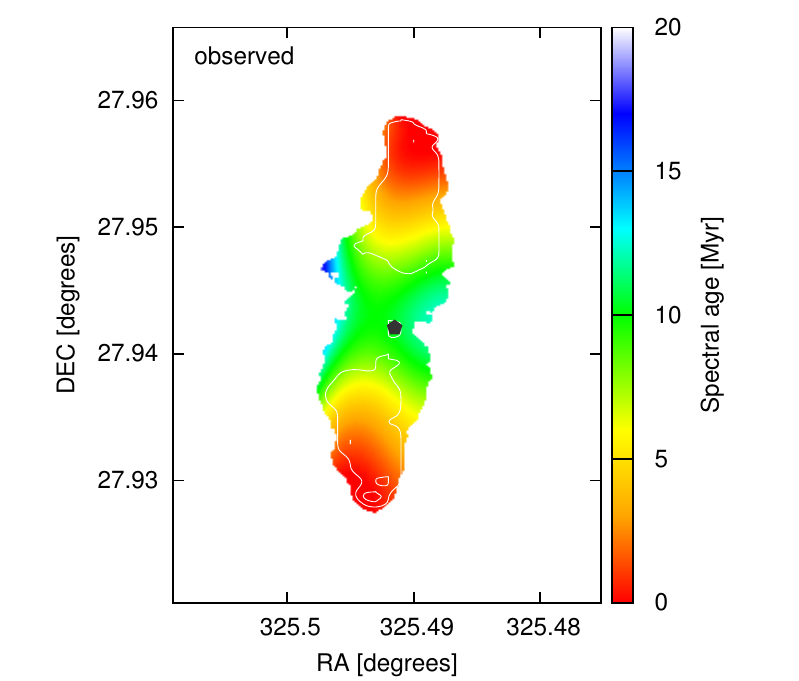} \!\!\!\!\!\!\!\!\!\!\!\!\!\!\!\!\!\!\!
\includegraphics[width=0.36\textwidth]{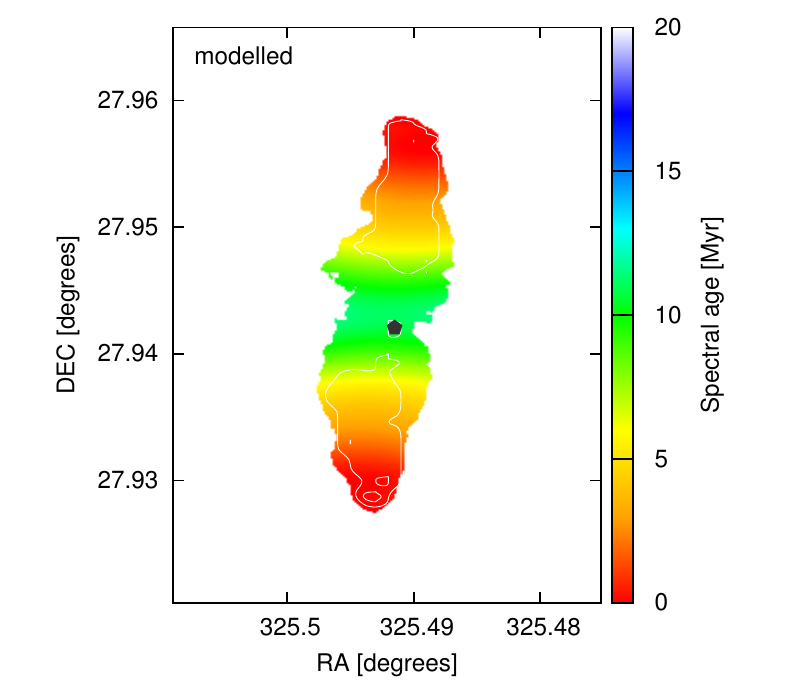} \!\!\!\!\!\!\!\!\!\!\!\!\!\!\!\!\!\!\!
\includegraphics[width=0.36\textwidth]{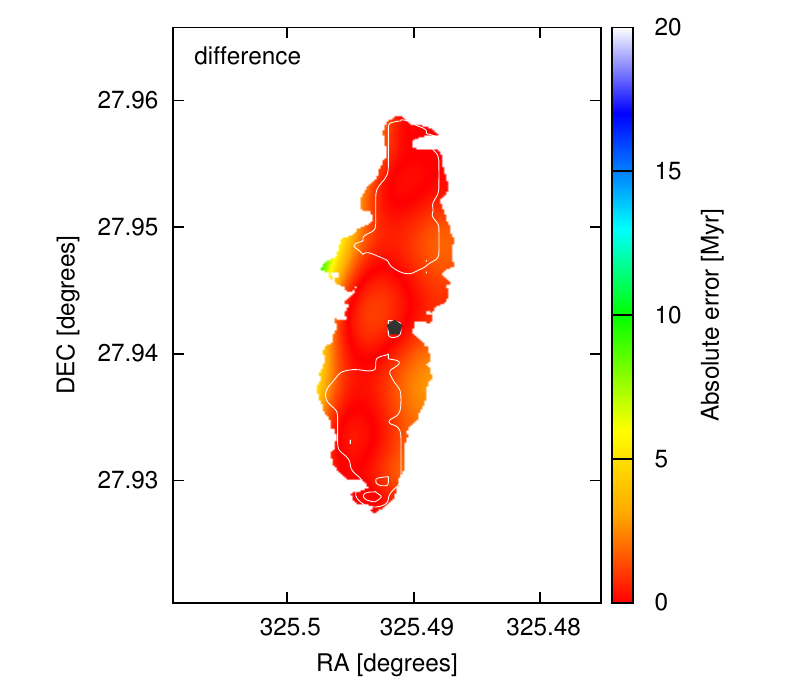}
\end{center}
\caption{Modelled spectral age distribution (centre) for the lobes of 3C436 calculated using the lobed FR-II/I dynamical model and resolved emissivity model; the size, axis ratio and total luminosity is constrained by observations (left). The absolute error between the modelled and measured spectral age distribution is shown in the right plot using the same colour scheme. Overlaid on each plot are $8.4\rm\, GHz$ radio contours showing that the older regions are also the faintest.}
\label{fig:3Cspec}
\end{figure*}

The multi-frequency surface brightness of 3C436 is calculated using the resolved emissivity model for the best fit jet power, source age and magnetic field strength. The radio SED arising from the synchrotron-emitting electrons at a given distance from the hotspot is fitted using the JP model. The spectral curvature of the emission arising from different aged regions of the lobes can therefore be quantified using the break frequency. The spectral age of the electrons at a given location is then calculated based on the field strength of the lobe using Equation 6 of \citet{Turner+2016}.
The spectral age distribution derived for 3C436 using our model is shown in the middle panel of Figure \ref{fig:3Cspec}. The measured ages \citep{Harwood+2013} are shown in the left panel and are corrected for the field strength fitted using our Bayesian algorithm; the original ages of \citet{Harwood+2013} were calculated assuming an equipartition magnetic field. Finally, the right panel of Figure \ref{fig:3Cspec} plots the absolute difference between the measured and modelled ages, showing there is at most a $2\rm\, Myr$ difference across the radio lobe, except for some few more isolated regions along the source edge which are likely not as well-mixed as predicted by our analytical model. This discrepancy may in part be due to edge effects in combining the observed maps at different spatial resolutions.

\begin{figure}
\begin{center}
\includegraphics[width=0.48\textwidth]{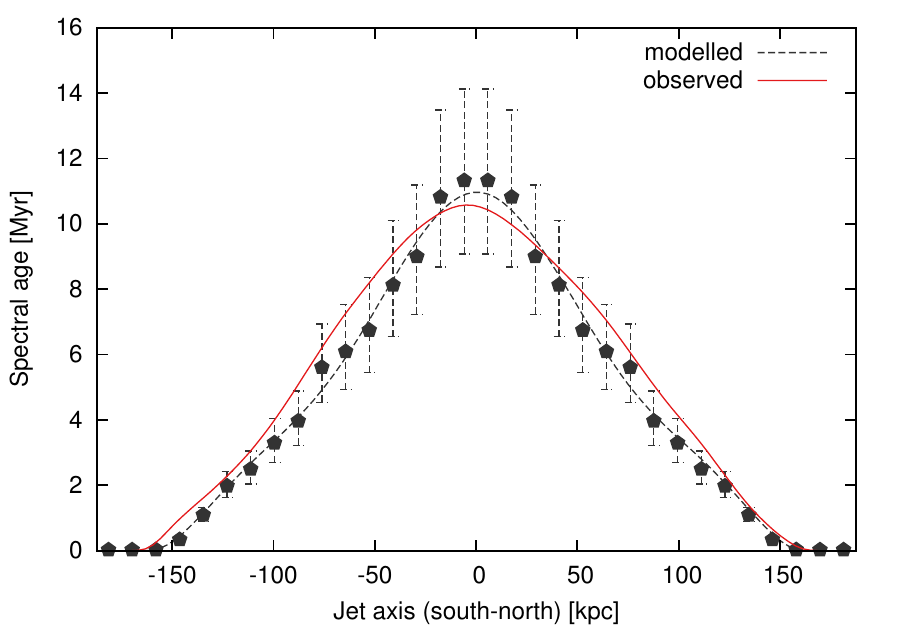}
\end{center}
\caption{Modelled (dashed black) and measured (solid red) spectral age along the jet axis of 3C436; i.e. the space curve passing through the central engine and otherwise centred on the circular lobe cross-section. The black pentagon show the modelled spectral age with $1\sigma$ uncertainties in the fitting the radio SED, including those from the Bayesian fit. The measured spectral ages have comparable uncertainties in their fits. The modelled spectral age along the centre of the lobes well matches the near-linear rise in age away from the hotspot observed by \citet{Harwood+2013}.}
\label{fig:3C436jetaxis}
\end{figure}

The modelled and observed spectral age distributions are further compared in Figure \ref{fig:3C436jetaxis}. The spectral age is measured along the jet axis to enable a more direct comparison; the jet axis is defined as the space curve which passes through the central engine and elsewhere is centred on the circular cross-section. The modelled spectral age along the centre of the lobes shows an approximately linear increase in age away from the hotspot consistent with the observations of \citet{Harwood+2013}. The modelled spectral ages agree with observations along the entire length of the jet. Both the observed and modelled spectral ages are only a small fraction of the source dynamical age close to the hotspots, but increase up to approximately one-third the fitted dynamical age near the core. {The general agreement between the two spectral age distributions suggests that our electron flow model is a reasonable approximation to the physical situation, and that neither inhomogeneities in the magnetic field nor the reacceleration of electrons are particularly important. Further investigation of these processes is deferred to future work.}

The discrepancy between the spectral and dynamical ages here is likely due to the invalidity of the assumptions used by the impulsive injection models, i.e. not all electrons in each slice have the same age. In particular, the hydrodynamical simulations shown in Figure \ref{fig:injectprofile} predict electrons injected as recently as $t_{\rm i} = 0.6t$ to be present in regions of the lobe closest to the core at the $1\sigma$ level. These younger electrons will extend the break in the radio SED out to higher frequencies; the JP and KP models largely fit the frequency above which no emission occurs yielding a peak age estimate of around $(1 - 0.6)t = 0.4t$. That is, these impulsive injection models tend to fit the spectral age of the youngest electrons that are contributing significantly to the emissivity of the radio SED. 

By contrast, applying the continuous injection model to the radio SED for the entire lobe reproduces the source dynamical age within uncertainties; this is expected since the CI model actually describes the physical situation, namely a region of homogeneous magnetic field strength with a continuous supply of freshly injected elections \citep[discussion in][]{Turner+2016}. Caution should be taken when measuring the spectral age of resolved radio sources; the measured age does not relate to the maximum or even average age of the electrons contributing to the SED.

\section{LOW-POWER RADIO AGN}
\label{sec:FLARING RADIO SOURCE MODEL}

The radio AGN counts at $z<1$ are dominated by faint low-powered FR-I class objects \citep{Willman+2008}, which will only become more prevalent in future surveys as sensitivity continues to increase. 
It is therefore vital to not only understand the powerful FR-II objects comprising much of the existing surveys, but also develop techniques to model the ubiquitous FR-Is. 
The definition of an FR-I radio source is not physically based, and simply specifies that the location of the surface brightness within the lobes is concentrated towards the core, i.e. brightest region lies less than one-half the lobe length from the core \citep{FR+1974}. Some FR-Is are observed to behave like FR-IIs without hotspots (the fat doubles), while others are twin jets which never inflate a lobe but instead expand out until they are disrupted by interaction with the intra-cluster medium. The dynamics of lobed FR-Is is similar to the FR-IIs, with the exception of the sites of electron acceleration. In this work, we will examine the flaring jet FR-Is, the archetypal example being 3C31 \citep[e.g. 3C31][]{Leahy+2013}. {The synchrotron-emitting electrons in these sources are assumed to undergo a single instance of acceleration at the jet flare point.}
The dynamics of the flaring jet FR-I radio plume as a whole are modelled using the subsonic expansion limit of the \citet{Turner+2015} model; however, the detailed dynamics of electron packets as they travel through the lobes remain to be described.

\subsection{FR-I formation and evolution}

Both the formation of the jet flaring point and the internal lobe plasma dynamics need to be characterised for FR-Is so that their spatial emissivity distribution can be determined. In this section, the dynamics of the lobe evolution and the bulk flow of lobe plasma is investigated using hydrodynamical simulations; this work is used to verify the results of our FR-I flaring jet dynamical model.

\subsubsection{Synchrotron-emitting electrons and lobe dynamics}

The large-scale radio source morphology depends crucially upon whether the jet maintains forward thrust during its interaction with the host environment.
If the jet does run out of thrust a flare point will form, resulting in the formation a type-I radio source. This can happen through one of two mechanisms: (1) the jet has a large opening angle causing it to run out of thrust before a recollimation shock forms \citep[see discussion in][]{Krause+2012}, or (2) the jet expands into an environment with a rising temperature profile \citep[e.g. observations in][]{Vikhlinin+2006} leading to a flare point even after the jet has collimated \citep[see simulations of][]{Massaglia+2016}.

In the first case, the ram pressure of the jet plasma reduces as it widens ($p_{\rm j} \propto 1/r^2$ for a conical jet) until it equals that of the ambient medium $p_{\rm x}$. The jet material will generate a recollimation shock upstream of this point when the sidewards ram pressure $\rho_{\rm j} v_{\rm j}^2 \sin^2 \theta$ and the external medium equilibrate.
This recollimation shock stabilises the jet if it reaches the jet axis before the progress of the jet stalls. In flat environments, analytical models \citep{Alexander+2006} and numerical simulations \citep{Krause+2012} show that the recollimation shock in jets with half-opening angles $\theta > 25^\circ$ occurs so far downstream that the terminal shock is unable to reach it and an FR-I flaring region forms \citep{Krause+2012}. The plasma downstream of the flare point streams transonically towards the maximum extent of the lobe. The lobes produced in this manner are very turbulent and have significantly elevated gas density and pressure at their ends. The simulated flare points well match those of typical astrophysical FR-Is \citep[e.g. 3C31;][]{Leahy+2013}, however this approach is less capable of modelling their radio plumes.

\citet{Massaglia+2016} investigated the formation of FR-I radio sources through the disruption of a collimated jet by a rising temperature profile. Their hydrodynamical simulations assume a standard falling density profile and prescribe the corresponding temperature profile which yields a constant pressure environment. The jet is disrupted in this case when a small pressure enhancement forms downstream (i.e. $p_{\rm j} \lesssim p_{\rm x}$). The plasma streams outwards with an approximately laminar flow from the flare point, leading to much less turbulence in the lobe than in the simulations of \citet{Krause+2012}. Both the simulated flare points and the radio lobes are consistent with observed radio sources. The initially conical jets of these sources may not be collimated on the small distance scales involved in these objects, especially for larger opening angles. The flare points in astrophysical FR-Is are likely formed through a combination of the techniques explored by \citet{Krause+2012} and \citet{Massaglia+2016}. 

{The bulk flow of plasma simulated in the low-power radio sources of \citet[][e.g. reference simulation in their Figure 4]{Massaglia+2016} is directed outwards from the core throughout almost the entirety of the radio plume. Entrainment between the plume and the ambient medium leads to a small region with a receding velocity; this region contains few synchrotron-emitting electrons and in any case has a near-zero negative velocity. In this scenario, the packets of synchrotron-emitting electrons accelerated at the flare point are expected to flow outwards indefinitely. Further, the injected electrons are expected to be approximately layered in the plume, with the newly accelerated particles closest to the core and the oldest electrons furthest from the central nucleus. The hydrodynamical simulations show that even the fastest electrons travelling along the jet axis take at least one-quarter of the source age to reach the end of the plume. However, the vast majority of electron population is in slower moving plasma away from the axis taking close to the source age to traverse this distance. 
The plasma towards the end of the plume will therefore be strongly skewed towards the oldest electrons in the population. The flow rate of the oldest synchrotron-emitting electrons is therefore simply the instantaneous expansion rate of the radio source. 
The pressure profile within the plume of a flaring jet FR-I should stabilise to the downstream external profile encountered by these initial synchrotron-emitting electrons; the internal sound speed must be low noting that the material in the plume originates approximately evenly from both the jet and the dense ambient medium \citep[comparison between FR-Is and -IIs provided in][]{Massaglia+2016}. 
We therefore assume all subsequent electrons flow forwards at the same rate as a function of distance from the core.}

\subsubsection{Flaring jet dynamical model}

The starting point of our dynamical model for twin-jet FR-I sources is the analytical model of \citet{Turner+2015}. Those authors modelled the evolutionary history of initially supersonic lobed radio sources with an ellipsoidal geometry. Here, we treat the jets as spherical cones. The jets initially expand supersonically, but upon forming a lobe quickly transition to subsonic pressure-limited expansion \citep[e.g.][]{Luo+2010}.
The radio lobe of the \citet{Turner+2015} dynamical model is an ensemble of small angular volume elements in pressure equilibrium. Each element of fixed angular width is assumed to receive an unchanging fraction of the jet power as the lobe expands to ensure the lobe maintains its initial shape. 
The lobe in the flaring jet model is a spherical sector and it is therefore expected that each element will expand identically. 
The angular volume elements are instead used to define which angles comprise the lobe and those which lie beyond the jet half-opening angle. The conical shape of the flaring jet lobe is therefore modelled by setting the lobe length $R(\theta) = 0$ for $\theta > \chi$, where $\chi$ is the half-opening angle of the lobe beyond the flaring point.

The luminosity of low-powered radio sources resembling a flaring jet is spatially much more complicated than that of the high-powered radio sources so far examined. Most notably, the magnetic field cannot be assumed to be homogeneous (on large scales) throughout the radio source due to the steeply falling pressure profile. 
{Further, \citet{Croston+2014} found that twin-jet FR-Is are more entrained further along the jets.
We assume here that the synchrotron plasma downstream of the flare point is well mixed with entrained ambient gas, so that the emitting electrons are effectively distributed across the prescribed cone cross section. This situation is is expected to be approached for the large-scale radio plumes modelled in this work.}

\subsection{Luminosity from synchrotron-emitting electrons}

{The synchrotron emissivity in flaring jet radio plumes is calculated in the same manner for both resolved and unresolved sources; the synchrotron-emitting electrons are accelerated at the flare point and flow outwards at the rate described by the dynamical model to their present-time location with ambient pressure.}

\subsubsection{FR-I surface brightness distribution}

\begin{figure}
\begin{center}
\includegraphics[width=0.485\textwidth]{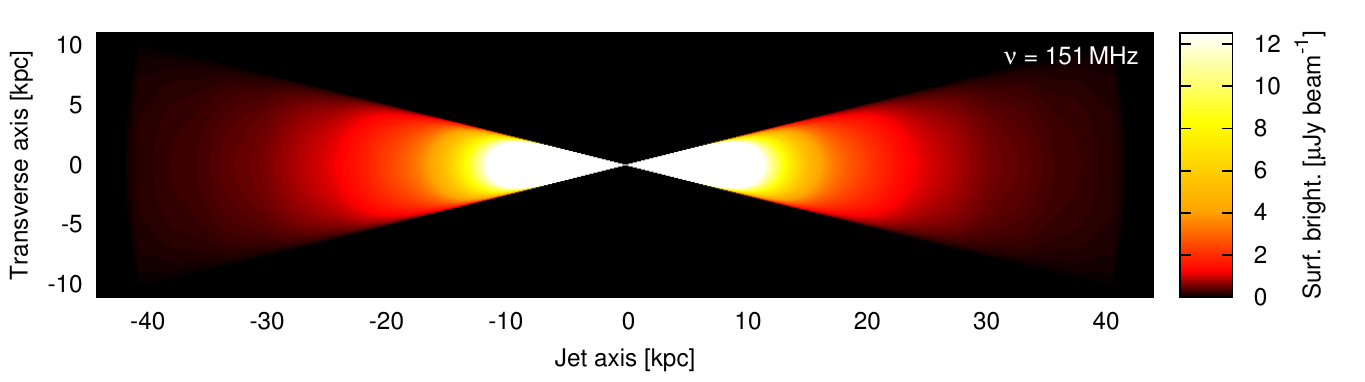} \\
\includegraphics[width=0.485\textwidth]{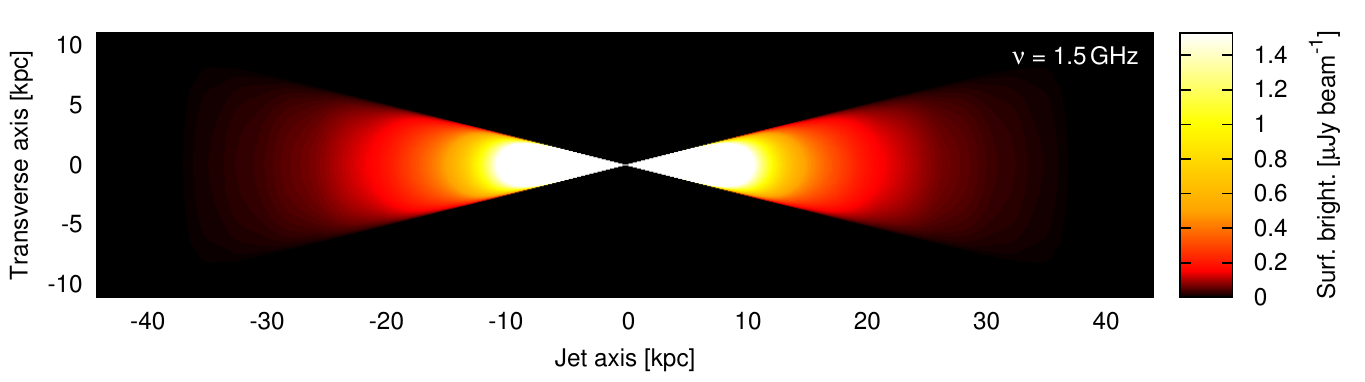} \\
\includegraphics[width=0.485\textwidth]{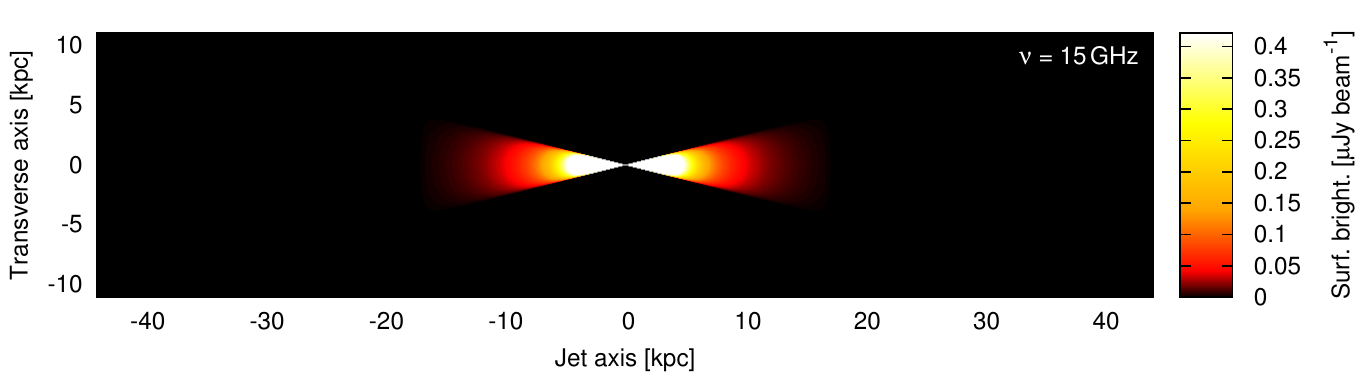}
\end{center}
\caption{Radio source emissivity as a function of lobe position including all loss mechanisms for three typical observing frequencies. The top plot shows $151\rm\,MHz$ radio emission, the middle plot is for $1.5\rm\,GHz$, and the bottom plot shows $15\rm\,GHz$ emission. The modelled source has a jet power of $Q = 10^{36} \rm\, W$, age of $t = 100\rm\,Myrs$, inhabits a $10^{14} \rm\, M_\odot$ halo mass environment at $z = 0.03$, expands with an plume half-opening angle of $\chi = 15^\circ$ and has an electron energy injection index of $s = 2.4$. The axes are scaled to the dimensions of the physical source size, i.e. the entire lobe does not emit radiation at any of these frequencies. The colour-bar scale is cropped to provide more contrast in the lobe at the expense of excluding the brightest emission near the core.}
\label{fig:FR1spatial}
\end{figure}

We calculate two-dimensional brightness distributions for flaring jet radio sources following the method described in Section \ref{sec:FR-II SPATIAL LUMINOSITY EVOLUTION}. 
The surface brightness distribution modelled for a typical flaring jet type FR-I is shown in Figure \ref{fig:FR1spatial} for frequencies of $151\rm\, MHz$, $1.5\rm\, GHz$ and $15\rm\, GHz$. As in the case of FR-IIs (Section \ref{sec:Two-dimensional emissivity}), the frequency dependent nature of the loss mechanisms reduces the duration over which the electron population emits synchrotron radiation at higher frequencies. However, unlike the powerful FR-II sources the acceleration site is close to the core with the electrons propagating outwards. The linear size out to which synchrotron emitting electrons reach may therefore be different at each frequency, with larger observable emitting regions expected at low frequencies where the losses are less severe.
The size is thus poorly defined for FR-Is, and will depend on both observing frequency and sensitivity; this is in marked contrast to the FR-IIs, where the measured length is unaffected as long as hotspots are visible.
{The observed linear size of the flaring jet sources is also found to reduce both with increasing redshift and source age, in a similar manner to the width of high-power FR-IIs.}
The correct measurement of the source size is vital to correctly estimate the source age, and thus total energy output of the AGN into its surroundings \citep{Turner+2015}. The corresponding lobe volume must be correctly measured to properly understand the suppression of cooling resulting from pushing gas to greater cluster-centric radii with longer cooling times \citep[e.g.][]{Cavagnolo+2010, Raouf+2017}.

\subsubsection{FR-I unresolved luminosity model}

\begin{figure}
\begin{center}
\includegraphics[width=0.48\textwidth]{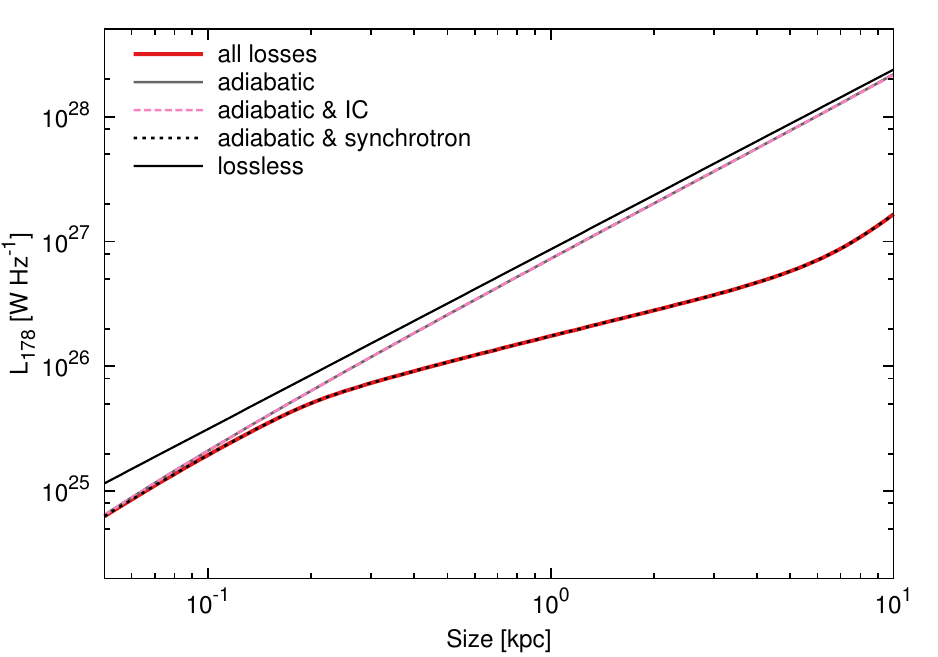} 
\end{center}
\caption{Luminosity-size tracks for an FR-I radio source at $z = 0.5$ with jet power of $1.3 \times 10^{35} \rm\, W$, expanding into an X-ray observed environment \citep{Vikhlinin+2006} scaled to a density of $\rho = 7.2\times 10^{-22} \rm\, kg\,m^{-3}$ at radius $2 \rm\, kpc$. Five profiles are plotted, showing the luminosity in the lossless case (solid black), with only adiabatic losses (solid grey), with adiabatic and synchrotron losses (dashed black), with adiabatic and inverse-Compton losses (dashed pink), and with all loss mechanisms (solid red).}
\label{fig:LDtrack}
\end{figure}

The luminosity calculation for unresolved flaring jet FR-I radio sources is much more involved than their high-powered FR-II counterparts. In particular, the pressure gradient down the flaring jet necessitates the use of the full resolved emissivity model, with the total luminosity derived by integrating the emissivity across all volume elements. 
The algorithm can be improved for the unresolved case since the location of the electron packets is unimportant; each electron age is directly mapped to a present-time magnetic field strength (i.e. electrons of different ages coast down the pressure gradient independently), resulting in a model as computationally efficient as for unresolved FR-IIs. 
The size-luminosity track for a typical FR-I with jet power $Q = 1.3\times 10^{35} \rm\, W$ is shown in Figure \ref{fig:LDtrack}. 
In the absence of radiative losses, the luminosity rises throughout the source lifetime as progressively more electrons are added to the high-pressure core.
{The size-luminosity track is largely unaffected by the inclusion of all but the synchrotron loss mechanism; the other loss processes only reduce the emission of the electrons found furthest from the core where the contribution to the total luminosity was already small due to the lower pressures.}  
This monotonic increase occurs because freshly injected plasma continually replaces the older material near the bright constant pressure core; the displaced plasma expands the plume and increases the overall luminosity. 

\subsection{Modelling spatial emissivity of real sources}
\label{sec:Modelling spatial emissivity of real sources}

In this section, we test our analytical spatial emissivity model by attempting to reproduce the surface brightness distribution of the typical FR-I source 3C31 \citep{Leahy+2013}. The intrinsic properties of 3C31 are fitted using the Bayesian framework described by \citet{Turner+2016}, based upon the observed source size, luminosity and break frequency arising from the oldest electrons, in addition to properties of the host environment. \citet{Heesen+2014} combined high-resolution observations of 3C31 from the \emph{Low Frequency Array} (LOFAR), the \emph{Very Large Array} (VLA) and \emph{Giant Metrewave Radio Telescope} (GMRT) to produce radio SEDs as a function of position within the radio lobe. We fit the JP model to their observed spectra for the furthest regions of the lobe to obtain estimates for the spectral break frequency. The properties of NGC\,383, the galaxy group hosting 3C31, are taken from \citet{Komossa+1999}. The shape of the pressure profile, gas mass and virial temperature for NGC\,383 are used here. We derive a jet power for both jets of $10^{37.2}\rm\, W$, source age of $200\rm\, Myrs$ and an equipartition factor of 0.31, consistent with the range found by \citet{Croston+2008}.

\begin{figure*}
\begin{center}
\includegraphics[width=0.36\textwidth]{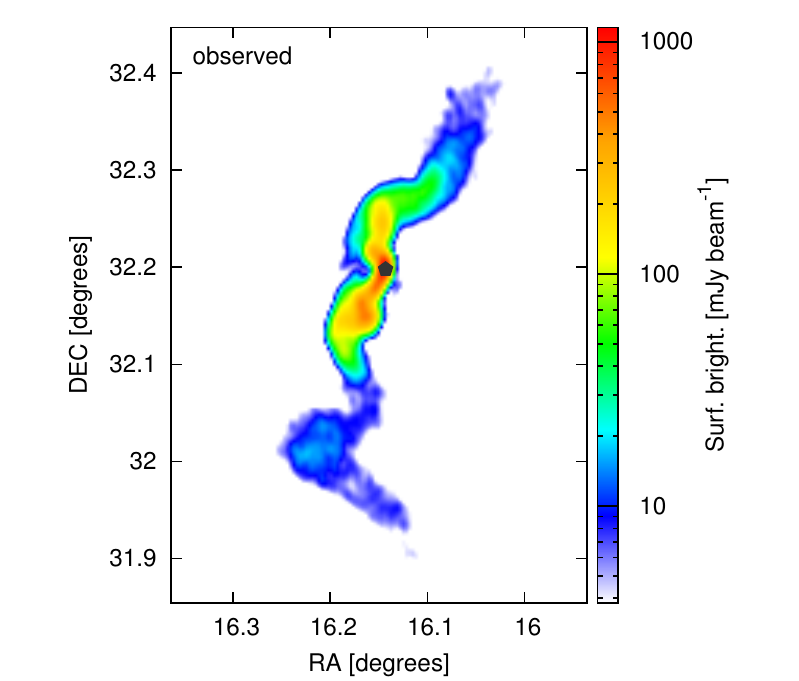} \!\!\!\!\!\!\!\!\!\!\!\!\!\!\!\!\!\!\!
\includegraphics[width=0.36\textwidth]{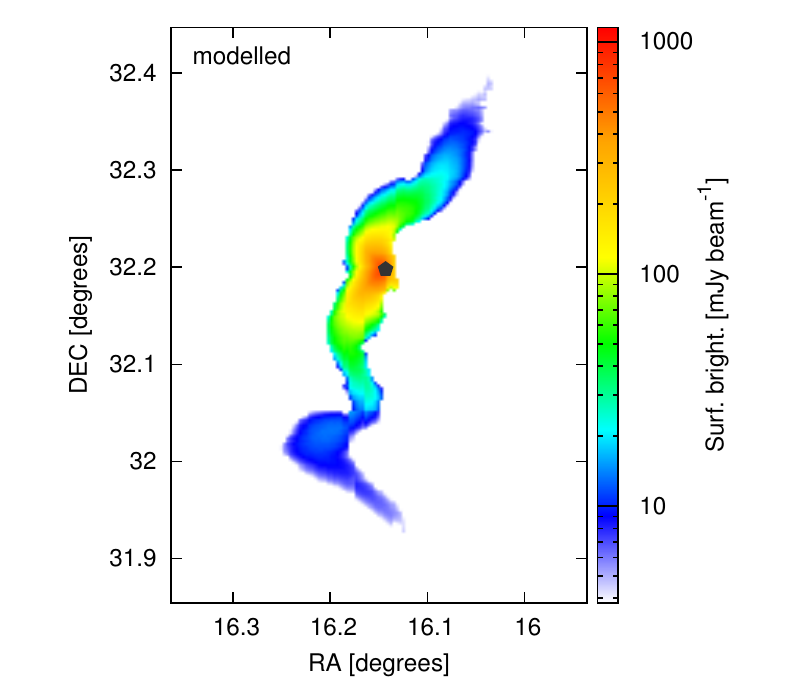} \!\!\!\!\!\!\!\!\!\!\!\!\!\!\!\!\!\!\!
\includegraphics[width=0.36\textwidth]{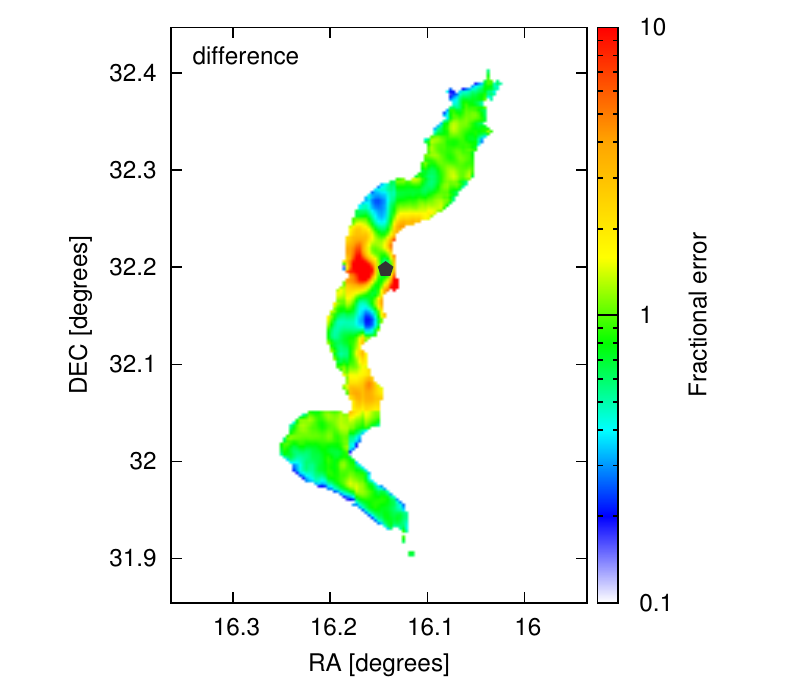}
\end{center}
\caption{Modelled $608\rm\, MHz$ radio emission (centre) from the lobes of 3C31 calculated using the flaring jet dynamical model and resolved emissivity model; the shape of the lobes and total luminosity are constrained by observations (left). The fractional error between the modelled and observed luminosity distribution is shown in the right plot. The modelled results agree everywhere except around the core; our modelling does not consider the early jet before the flaring point leading to underestimated emission along parts of the jet and overestimated emission off axis.}
\label{fig:3C31sim}
\end{figure*}

The resolved emissivity model (Section \ref{sec:SPATIALLY RESOLVED LOBE LOSSES}) is used to predict the emissivity from the electrons of various ages in 3C31 for the best fit jet power, source age and magnetic field strength. These electrons are distributed within the observed volume occupied by 3C31 \citep[i.e. an approximately helical shape;][]{Leahy+2013}; the pressure-limited expansion model we assume from the growth of flaring jet radio sources dictates the radial location of the electrons of different ages. The modelled radio source emissivity for 3C31 calculated at $608\rm\, MHz$ is plotted in Figure \ref{fig:3C31sim}. The pressure-gradient, loss processes and $1/r^2$ expansion lead to a rapid decrease in surface brightness away from the central engine, i.e. $S \propto r^{-\beta - \xi - 1}$ where $\beta$ and $\xi$ are the exponents of the density and temperature profile, and one dimension of the $1/r^2$ volume expansion is assumed to lie along the line-of-sight. Further, the three-dimensional nature of the radio source leads to decreased emission from line-of-sights passing through the edge of the source (i.e. less plasma here than towards the centre), as seen in observations of 3C31. This effect also results in the region of enhanced brightness in the westwards kink in the tail of 3C31. 
The observed emissivity of 3C31 is well characterised by the combined FR-I dynamical and luminosity model, as shown in the right panel of Figure \ref{fig:3C31sim} which plots the fractional error between the modelled and observed emissivity distributions.
However, the spatial emissivity model does not accurately reproduce every detail of 3C31. In particular, the region around the central engine is unrealistically modelled assuming isotropic jets. As a result, our model greatly overestimates the emission off the jet axis and slightly underestimates the brightness along the jet. This can largely be corrected by assuming that constant opening angle jets feed into the lobes of 3C31, though we do not pursue this further since we aim solely to validate our analytical model at this time.

\begin{figure}
\begin{center}
\includegraphics[width=0.48\textwidth]{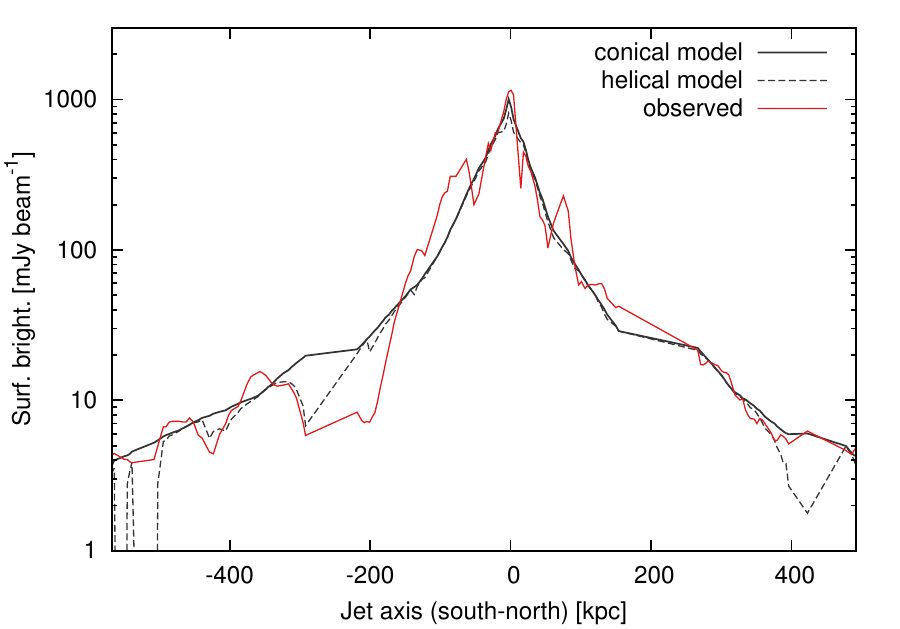}
\end{center}
\caption{Modelled and observed (solid red) $608\rm\, MHz$ radio emission along the jet axis of 3C31; i.e. the space curve passing through the central engine and otherwise centred on the circular lobe cross-section. {The modelled surface brightness is shown assuming either a simple conical cross-section (solid black) or a more complicated helical structure (dashed black) based on the observed width of the plume.} The modelled emissivity along the centre of the plumes matches the overall shape of the observed profile well. Further, {the helical model reproduces} some of the finer details such as the peak near the central engine, and the spikes in luminosity due to the bends in the lobe at around 300 and 450 kpc south of the core.}
\label{fig:3C31jetaxis}
\end{figure}

The similarity of our modelling and the observed surface brightness distribution is further examined in Figure \ref{fig:3C31jetaxis}. The surface brightness is measured along the jet axis to enable a more direct comparison; the jet axis is defined as the space curve following the path of the jets where visible near the core or the centre of the circular lobe cross-section at larger radii. The overall shape of the modelled surface brightness along this jet axis is very close to that of the observed profile. The peak near the central engine and the spikes in emissivity due to the bends in the lobe at around 300 and $450\rm\,kpc$ south of the core are also reproduced by our modelling. The higher than modelled surface brightnesses approximately $100\rm\, kpc$ from the core result from the jets, whilst the lower than expected surface brightness approximately $200\rm\, kpc$ south of the nucleus is likely due to either an unmodelled feature in the environment or any cork-screwing of the jet being poorly captured by our model.
{This comparison is also made with a simpler model assuming a conical cross-section (see Figure \ref{fig:3C31jetaxis}). This model successfully matches the general shape of the observed emissivity profile but cannot reproduce the finer details. The surface brightness in the plumes of low-powered radio sources can therefore be well described by considering only the reduced electron density and magnetic field strength along the length of the source. In particular, this agreement is possible despite our model ignoring entrainment, turbulent electron reacceleration, and any small-scale structure in the magnetic field.}

\subsection{Flaring jet FR-I spectral classification}
\label{sec:Flaring jet FR-I spectral classification}

The radio SED measured for radio sources depends crucially on the sky area over which the flux density is integrated to generate the spectrum. The difficulty in measuring the linear size of flaring jet FR-I radio sources may lead to significant emission from the radio lobes being excluded if it lies below the noise floor of the survey. The weak synchrotron emission from the large volume surrounding the detected regions of the lobe should be spatially coherent and may be recoverable; this of course assumes such consideration is taken during observations.

\begin{figure}
\begin{center}
\includegraphics[width=0.48\textwidth]{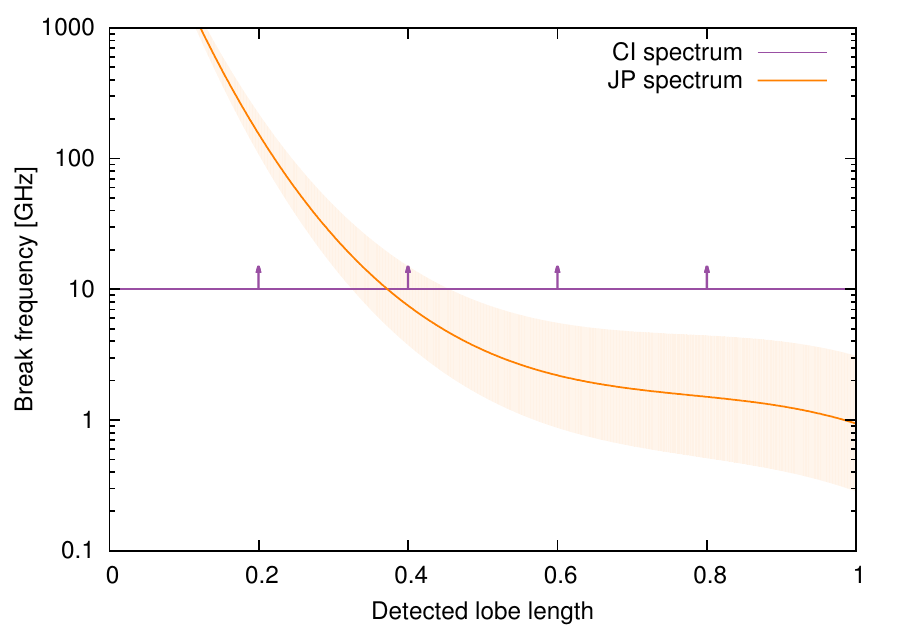}
\end{center}
\caption{Break frequency fitted to the radio SED from strips $[r - dr, r + dr)$ along a typical flaring jet radio source (parameters as in Figure \ref{fig:FR1spatial}). The shading plots the $1\sigma$ uncertainties in the JP model fits of the spectral break. The lower limit for the continuous injection (CI) model fit to the spectrum is also included; the break frequency cannot be fitted to the unresolved spectrum since the inhomogeneous magnetic field violates a model assumption.}
\label{fig:flarebreak}
\end{figure}

The radio SED arising from an unresolved flaring jet FR-I source is not well modelled by either the standard impulsive (e.g. JP and KP) or continuous injection models. In particular, these sources violate the assumption of a homogeneous magnetic field leading to a flatter SED\footnote{The radiative losses from the oldest synchrotron-emitting electrons lead to the steepening of the radio SED at high-frequencies. However, in flaring jet FR-Is these electrons are located furthest from the core where the magnetic field strength is weakest, and thus contribute less to the total luminosity than expected under the assumption of a homogeneous magnetic field.} beyond the break frequency than expected. Any spectral break observed in the SED therefore can not be related to the age of the source.
Moreover, the location of the break is largely determined by the bright, freshly injected electrons near the core. 
The lower bound of the spectral break (fitted using only the start of the turnover) is determined to be greater than 10-$100\rm\, GHz$ for typical FR-Is at low-redshift ($z \ll 1$) with jet powers of $\sim 10^{36}\rm\,W\,Hz^{-1}$, sources ages less than a few hundred Myrs, and environments with halo mass $10^{12}$-$10^{15}\rm\,M_\odot$.
Spectral turnovers much above the highest observing frequency are unlikely to be detected and the radio source SED will therefore be assumed to have no curvature implying a very young source. Further, the host galaxy may significantly impact the low frequency emission ($<1\rm\, GHz$) near the core due to synchrotron self-absorption or free-free absorption. The spectrum arising from the bright inner regions of flaring jet radio sources may thus resemble that of gigahertz peaked spectrum (GPS) or compact steep spectrum (CSS) radio sources. 

The emission arising from narrow radial strips of the radio lobe can instead be examined, to avoid the radio SED being dominated by the young electrons near the acceleration site, and to satisfy the assumption of a homogeneous magnetic field. The spectral break frequency fitted to the similar age electrons (i.e. impulsive injection model) in strips at different radii from the central engine is shown in Figure \ref{fig:flarebreak}. The break frequency measured at the electron acceleration site is consistent with that fitted for the unresolved radio source, however the spectral break falls off rapidly with radius to approximately $1\rm\, GHz$ at about half the source size. The break frequency remains approximately constant at $1\rm\, GHz$ (for this typical FR-I) for the strips at larger radii. This is theoretically expected since the age of the electron packets increases with radius as $\tau \propto r^{3 - \beta}$ and the magnetic field as $B \propto r^{-\beta/2}$ \citep[from subsonic limit of][]{Turner+2015}, hence the break frequency $\nu_{\rm b} = \tau^{-2} B^{-3}$ becomes independent of radius in steep environments with $\beta \sim 1.7$ (i.e. those far from the core).
This result enables the spectral age of resolved flaring jet sources to be easily calculated. Following the method of \citet{Turner+2016} the spectral age can be used to constrain the intrinsic source age in our dynamical models.

The full extent of flaring jet radio sources may not be detected in objects viewed at close to the surface brightness detection limit. The possibility of extended emission surrounding the cores of these sources must be considered to properly understand the AGN population and their energetics. Integrating the surface brightness to well beyond the observed source size may be required to include the contribution from all synchrotron emitting electrons originating from the radio AGN. Any high-frequency curvature observed in an SED measured over such an extended region would suggest the source is not compact, but rather has faint, extended emission. The large FR0 population comprising GPS and CSS radio sources could at least partly be explained be explained by low jet power flaring jet FR-I type sources. The emission in such sources would be dominated by the young and bright electrons in the high emissivity core, whilst older electrons at larger distances would be too faint to be detected.

\section{UNDERSTANDING AGN FEEDBACK AND ENERGETICS}
\label{sec:UNDERSTANDING AGN FEEDBACK AND ENERGETICS}

The energy imparted on the AGN host galaxy and its larger-scale environment is largely driven by two mechanisms: (1) the energy input into the host environment directly from the jets (i.e. $E = Qt$), and (2) the suppression of cooling by pushing gas out to large galactocentric radii with longer cooling times. The calculation of the source active lifetime, and thus energy, using dynamical models requires accurate source size estimates \citep{Turner+2016}. The linear size of an FR-II radio source can be easily measured in any resolved source since the brightest regions are located at the extremities of the lobes. By contrast, the full extent of FR-Is with core-dominated emission may not be observed. Estimating the volume of gas cleared by the expanding radio lobes may be miscalculated in both morphological classes; FR-IIs may have their width underestimated whilst the length of FR-Is is likely to be miscalculated.

\subsection{Measuring lobe sizes and axis ratios}

Survey sensitivity may greatly affect the measurement of radio AGN energetics through the underestimation of either their width or length. In this section, a well-studied sample of 3C sources are used to quantify the degree to which the width and length are underestimated. The surface brightness of 3C sources is typically several orders of magnitude greater than the surface brightness limit of the observations; this ensures the full extent of the AGN is visible. The change in length, width and axis ratio of these sources can then be determined as a function of sensitivity to understand observations in future surveys.

Radio source FITS images for 3C sources with $z < 1$ are taken from the online database of \citet{Laing+1983}\footnote{\url{http://3crr.extragalactic.info}}. The lengths, widths and axis ratios of the radio source lobes are recalculated with various survey surface brightness sensitivity limits, in addition to a physical measurement based on the actual signal-to-noise of the image. The 3CRR survey selected bright and then unresolved objects with $178\rm\, MHz$ flux densities $>10.9\rm\, Jy$ \citep{Laing+1983}; these have since been observed at higher resolutions and with several orders of magnitude greater sensitivity. The FITS images are available from observations at different frequencies for many of the 3C sources, typically $1.4$, $4.9$ or $8.5 \rm\, GHz$. The brightnesses in each of these maps are arbitrarily scaled to a $1.5 \rm\, GHz$ observing frequency with a $1\rm\, arcsec^2$ beam to enable a cross-frequency comparison.

The processing of the FITS images to measure the shape of the radio lobes proceeds as follows. Material potentially belonging to the lobes is firstly identified by flagging out all emission below the $5\sigma$ noise level of the image. Connected regions of emission are identified; proceeding outwards from the optical host galaxy, these clumps of emissions are flagged as lobes until at least 90\% of the image flux density has been included in lobes. The lobe selections made using this technique are confirmed with a visual inspection.
The most distant two points (from each other) in the flagged region are identified as belonging to different lobes. The lobe length is measured as the distance from the core to the most distant pixel in each lobe, and the width taken as the greatest width across the lobe normal to this axis. The axis ratio is calculated as the ratio of the lobe length to its cross-sectional radius, the radius of the third of the lobe closest to the hotspot is not included. We make this decision to allow a fair comparison with our model which does not consider the enhanced emission near the overpressured hotspot. The morphology of the lobe is additionally characterised by taking the emissivity weighted mean length. That is,

\begin{equation}
F\!R = \frac{2 \sum s_{\rm i} x_{\rm i}}{\max\{x\}\sum s_{\rm i}} + 0.5 ,
\label{FRnumber}
\end{equation}

where $s_{\rm i}$ is the flux density from a cell distance $x_{\rm i}$ along the jet axis. Here an $F\!R$ number of 0.5 corresponds to a source with all emission at the core and a number of 2.5 to all emission at the hotspot. By definition, \citeauthor{FR+1974} type 1 sources have an $F\!R$ number less than 1.5 whilst FR-IIs comprise the remainder of the population \citep[c.f.][]{Krause+2012}. The observing frequency may affect the morphological classification determined for a particular source, however the effect of changing the surface brightness sensitivity limit on these classifications should be frequency independent.

\subsection{Sensitivity and Fanaroff-Riley morphology}

\begin{figure}
\begin{center}
\includegraphics[width=0.45\textwidth]{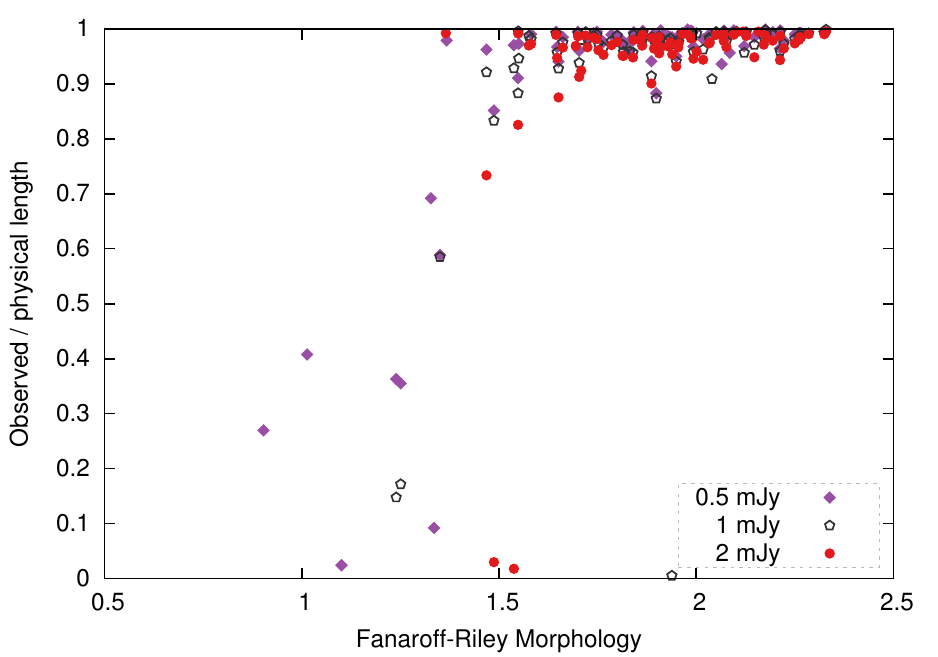}
\end{center}
\caption{Lengths of 3C radio sources detected for three survey sensitivities (per beam) as a function of the morphological classification. Radio lobes with an FR-II morphology (i.e. number greater than 1.5) have their lengths measured correctly irrespective of the survey sensitivity, whilst the length of lobes with emission concentrated towards the core is a strong function of sensitivity.}
\label{fig:FRlength}
\end{figure}

\begin{figure}
\begin{center}
\includegraphics[width=0.45\textwidth]{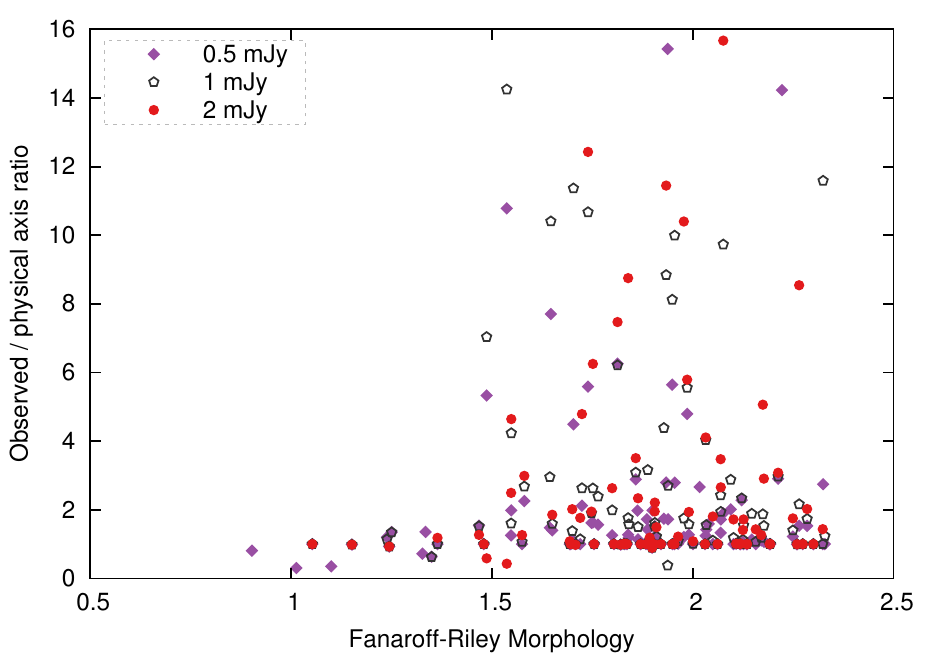}
\end{center}
\caption{Axis ratios of 3C radio sources detected for three survey sensitivities (per beam) as a function of the morphological classification. The radio lobes of FR-I sources tend to have their axis ratios correctly measured irrespective of the survey sensitivity, however the measured axis ratios of FR-II objects can vary drastically with sensitivity.}
\label{fig:FRaxis}
\end{figure}

Biases in the measurement of radio source properties due to observing with insufficient sensitivity are expected to depend on the \citeauthor{FR+1974} morphology. The fractional difference in the detected and physical length and axis ratio of the lobes of the 3C sources is investigated for different survey sensitivities in Figures \ref{fig:FRlength} and \ref{fig:FRaxis}. Radio lobes with an FR-II morphology have their lengths measured correctly irrespective of the survey sensitivity, whilst the length of lobes with emission concentrated towards the core is a strong function of sensitivity. This makes qualitative sense as the bright hotspots of FR-IIs should remain visible after any other region in the lobe, whereas in FR-Is the regions furthest from the core tend to be the faintest. Meanwhile, the lobes of FR-I sources tend to have their axis ratios correctly measured irrespective of sensitivity, however the measured axis ratios of type 2 objects can vary significantly with sensitivity. The axis ratios of type 1 objects are likely measured correctly despite their incorrect lengths because of their approximately conical shape. The axis ratios of FR-IIs will be incorrectly measured if emission from the lobe falls below the detection limit whilst that close to the hotspots remains detected.

\subsection{Observed axis ratio of powerful FR-II sources}

The axis ratio of lobed FR-IIs is expected to vary with sensitivity for one of three reasons: (1) only the hotspots may be detected leading to a very high axis ratio, (2) synchrotron loss mechanisms may reduce emission away from the hotspot so that the luminosity from the widest region near the core is greatly reduced, or (3) in weaker sources the circular cross-section of the lobe may cause the emissivity to reduce as $\sqrt{1 - y^2}$, where $y$ is the fractional distance from the jet axis to the edge of the lobe. We attempt to exclude the first of these reasons from our analysis by measuring the source width across only the two-thirds of the source closest to the core; if zero width is measured then we can assume only the hotspots are visible. Moreover, the size of the hotspot as a function of sensitivity has no bearing on the physical width of the lobe. By contrast, the other two mechanisms are included in our model enabling observations of radio sources taken near the detection limit to potentially have their axis ratio corrected.

\begin{figure}
\begin{center}
\includegraphics[width=0.48\textwidth]{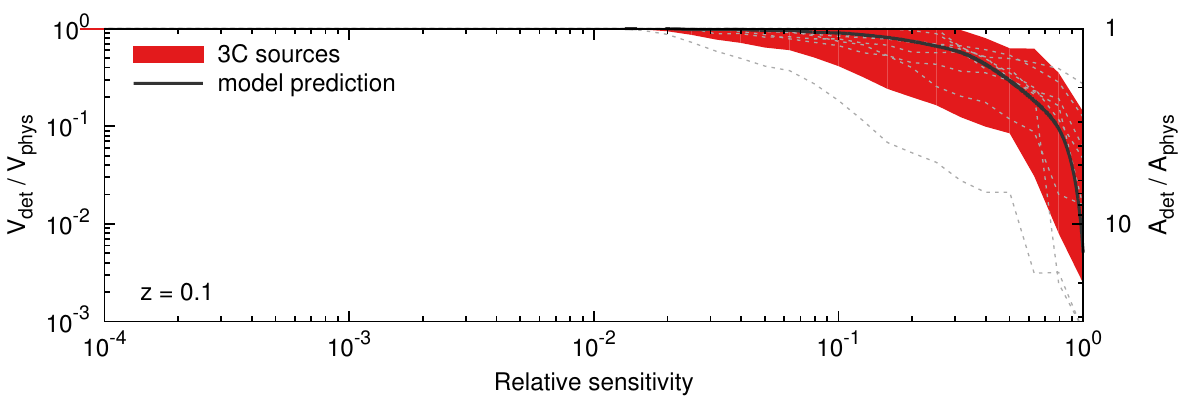}\\
\includegraphics[width=0.48\textwidth]{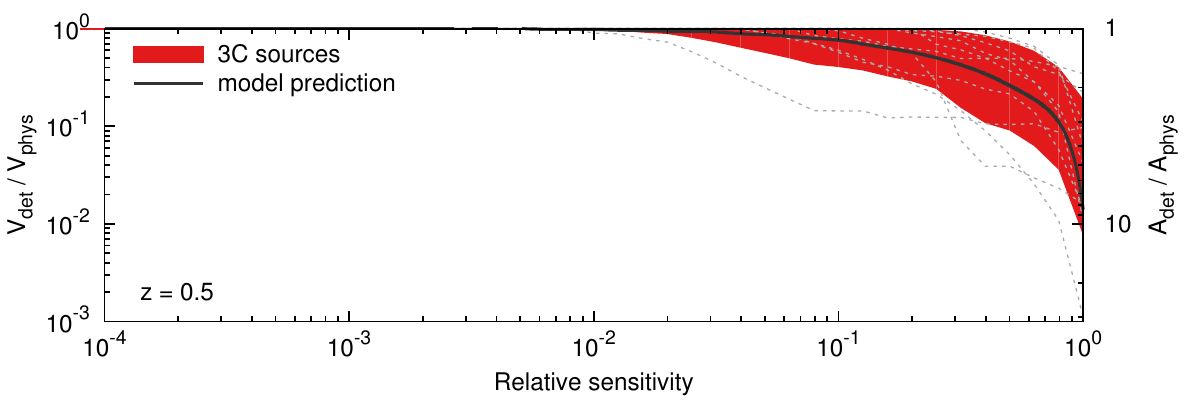}\\
\includegraphics[width=0.48\textwidth]{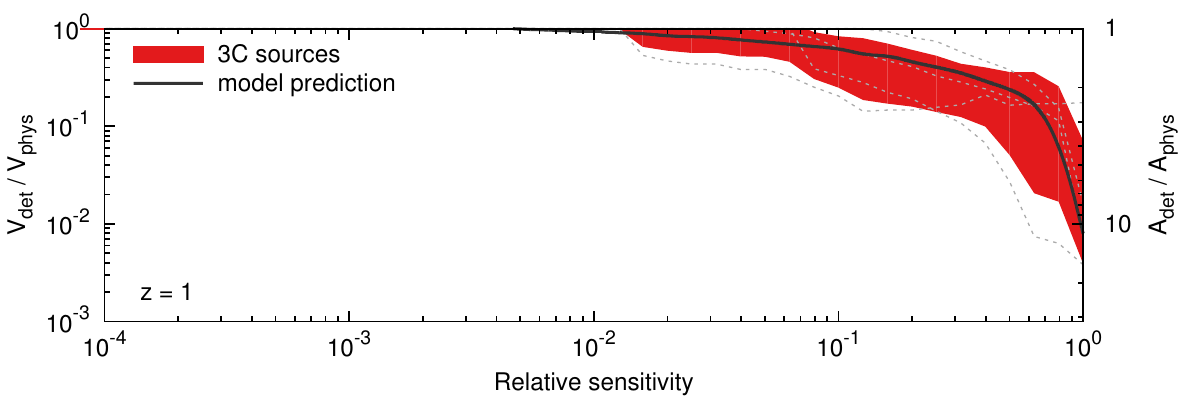}
\end{center}
\caption{Volume of lobed FR-II sources in the 3C survey detected as a function of the survey sensitivity binned by the host galaxy redshift. The left-hand vertical axis shows the fraction of the ``total'' lobe volume detectable by an arbitrary sized beam for various survey sensitivities; the sensitivity is scaled so that unity corresponds to the point when a given lobe first becomes detectable. The right-hand vertical axis similarly shows the fraction of the lobe width detectable, i.e. the detected axis ratio divided by the physical axis ratio. The grey dashed lines show the tracks of each 3C radio lobe in this phase-space, with the red shading indicating the $1\sigma$ spread in the distribution. The solid black line plots the theoretical expectation from our model, using the average redshift and age of the sources in the plot. This is again scaled so that a sensitivity of unity corresponds to the point at which the source first becomes detectable, and is additionally scaled so that with the entire volume is detected with infinite sensitivity. The 3C radio sources are grouped into three redshifts: $z \leqslant 0.3$ (top), $0.3 < z \leqslant 0.7$ (middle) and $z > 0.7$ (bottom).}
\label{fig:3Caxisratios}
\end{figure}

The observed axial ratios for FR-II radio sources in the 3C survey is investigated in Figure \ref{fig:3Caxisratios}. The \citeauthor{FR+1974} classification in Equation \ref{FRnumber} is used to robustly classify type 2 sources, with $F\!R < 1.75$ excluded due to their more core dominant nature and $F\!R > 2.25$ excluded due to the lack of lobe emission. The axis ratio of these sources is measured at various survey sensitivities and compared to the best estimate measure using the FITS image. The detected axis ratio of each source falls off steeply as the sensitivity decreases. We note that this slope depends on the inverse-Compton losses, and we therefore bin the 3C sources into three redshift bins. The sensitivity scale for each of the 3C sources is shifted such that unity corresponds to the point when a given lobe first becomes detectable. The shapes of the 3C source profiles are compared with each other and to the model prediction for a radio source typical of the 3C population in the redshift bin, as shown in Figure \ref{fig:3Caxisratios}. The theoretical model prediction for the measured axis ratio and lobe volume as a function of survey sensitivity is again found to be consistent with that of the observed 3C sources in each of the three redshift bins. 

The rate at which the detected lobe width reduces as observations become less sensitive is expected to be a function of the strength of the inverse-Compton loss mechanism. The oldest synchrotron-emitting electrons are most subject to inverse-Compton losses; these electrons are found furthest from the hotspot in the widest regions of the lobe closer to the core. The inverse-Compton loss mechanism is most prevalent at higher redshifts where the CMB radiation is stronger, and in older sources where it has longer to act on the electrons. Both the observed 3C sources and our model show that the axis ratio will be measured incorrectly for a greater range of sensitivities in higher redshift sources than at lower redshifts (see Figure \ref{fig:3Caxisratios}). Equivalently, we can say that the surface brightness distribution of low-redshift FR-II radio lobes is more homogeneous than in their high-redshift counterparts. 

We find that the detected lobe width will be accurately measured except when the brightest emission from the source is within a factor of a few of the survey detection limit. However, in a volume-limited survey radio sources are preferentially seen near the detection limit since lower luminosity objects are more ubiquitous. The detected lobe width and volume may therefore not reflect that of the physical radio source in much of the observed population.

\subsection{Visible linear size of flaring jet FR-Is}
\label{sec:Visible linear size of flaring jet FR-Is}

The surface brightness of flaring jet FR-I radio sources falls off rapidly away from the acceleration site near the AGN core, both due to the falling pressure profile and the radiative loss mechanisms. The sensitivity of the survey observing these sources may therefore lead to very different measurements of the source linear size. Further, the loss mechanisms reduce the energy of the electron distribution sufficiently that at some frequencies sections of the lobe cannot be observed irrespective of the sensitivity. When modelling 3C31 in Section \ref{sec:Modelling spatial emissivity of real sources} we found the surface brightness in flaring jet FR-Is falls off with radius from the core as $S \propto r^{-\beta - \xi - 1}$. The detected length of these sources is therefore expected to increase with improving survey sensitivity as $D \propto S^{-1/(\beta + \xi + 1)}$, though this simplistic formalism ignores the loss processes. For typical FR-I environments ($\beta \sim 1$ and $\xi \sim 0$) this relation simplifies to $D \propto S^{-0.5}$. 

\begin{figure}
\begin{center}
\includegraphics[width=0.48\textwidth]{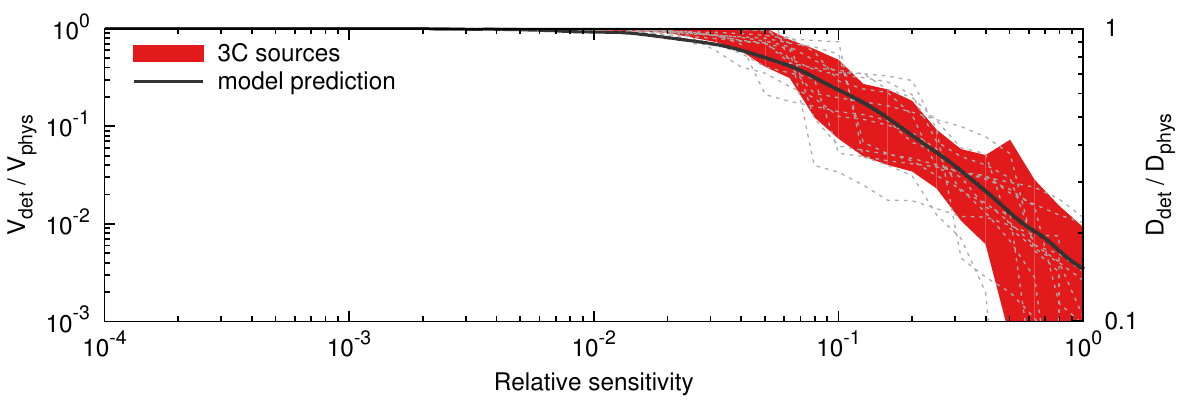}
\end{center}
\caption{Volume of flaring jet and core dominated radio sources in the 3C survey detected as a function of the survey sensitivity. The left-hand vertical axis shows the fraction of the ``total'' lobe volume detectable by a beam of arbitrary size for various survey sensitivities; the sensitivity scale for every source is shifted so that all the profiles lie on top of each other. The right-hand vertical axis similarly shows the fraction of the lobe length detectable. The grey dashed lines show the tracks of a each 3C radio lobe in this phase-space, with the red shading indicating the $1\sigma$ spread in the distribution. The solid black line plots the theoretical expectation from our model, using the average redshift and age of the sources in the plot.}
\label{fig:3Clengths}
\end{figure}

The volume and length detected for flaring jet and core dominated FR-I radio sources in the 3C survey is investigated in Figure \ref{fig:3Clengths}. The \citeauthor{FR+1974} classification in Equation \ref{FRnumber} is used to initially identify type 1 radio source lobes before being visually classified. The length of these sources is measured at various survey sensitivities and compared to the best estimate measured using the FITS image. The detected length of each of these FR-Is falls off approximately log-linearly as the sensitivity limit is raised and with similar slope, though there are sharp jumps in the detected length due to the presence of knots in the flow. The rate at which the detected lobe length reduces with sensitivity is expected to be approximately the same for all sources expanding into environments falling-off with similar power-laws; inverse-Compton losses also affect the maximum distance out to which emission occurs and slightly steepens the luminosity profile. The sensitivity scale for each of the 3C sources is therefore shifted such that the profile for each source is stacked on top each other. This enables a direct comparison of their shapes with each other and to our model prediction, as shown in Figure \ref{fig:3Clengths}. Our model is run for a radio source typical for the 3C population inhabiting a host halo environment of mass $10^{14} \rm\, M_\odot$ at redshift $z=0.5$. This theoretical prediction for the measured length as a function of survey sensitivity is found to be consistent with that of observed 3C sources.

Radio sources in deeper large-scale surveys such as the \emph{NRAO VLA Sky Survey} \citep[NVSS;][]{Condon+1998} and \emph{Faint Images of the Radio Sky at Twenty-Centimeters} \citep[FIRST;][]{Becker+1995}, in addition to the upcoming ASKAP \emph{Evolutionary Map of the Universe} survey \citep[EMU;][]{Norris+2011}, will predominately lie close to the detection limit. The weaker sources populating these surveys are expected to be predominately FR-Is \citep{Willman+2008}, or perhaps FR0s \citep{Baldi+2016} if these are completely separate populations. The measured length of most radio sources detected in these surveys is thus likely to be a lower limit on the true source size. Extended sources comprising multiple beams may still be detectable even if the flux density in each beam is below the survey detection limits, i.e. several beams with flux densities below the minimum sensitivity will have correlated ``noise'' and may yield a statistically significant detection. In general, simulating these flaring jet radio sources using our dynamical and resolved emissivity models should provide a means to constrain their energetics by fitting their observed surface brightness profile.

\subsection{Sensitivity limited feedback energetics}

The kinetic-mode of feedback is the main mechanism though which AGN impart energy into their host environments since $z = 1$; specifically the clearing out the hot gas at cluster centres through radio bubbles, or directly through shock heating in FR-IIs and the viscous dissipation of sound waves in FR-Is \citep{Bower+2006, Croton+2006, Croton+2016, Shabala+2009, Fanidakis+2011, Fabian+2012, Dubois+2013, Vogelsberger+2014, Schaye+2015}. The radio-only observations of the lobe volume are susceptible to surface brightness sensitivity limits, but in practice X-rays are generally used to directly measure the gas displaced from the cluster centre \citep[e.g.][]{Boehringer+1993, Fabian+2000, Fabian+2003, Fabian+2012}. However, the energy input directly through the FR-I/II jets is determined based upon the properties of the radio source, including its length and width; these are dependent on the sensitivity limit, potentially leading to a similar dependency for the energy.

The uncertainty in measuring the energy input by powerful FR-IIs directly into their environments through shock heating is quantified by fitting the jet powers, ages and magnetic field strengths of the 3C sources \citep[see][]{Turner+2016}. These parameters describing the AGN energetics are fitted for a range of axis ratios ($A = 2$, 4, 6 and 8) to enable their variation with the measured lobe width to be characterised. The jet power is found to scale with the axis ratio as $Q \propto A^{1.08\pm0.09}$ whilst the source age is inversely related as $t \propto A^{-1.07\pm0.02}$. As a result, the energy input directly by the jet into the host environment is independent of the axis ratio measurement within $1\sigma$ uncertainties. By contrast, the magnetic field strength is somewhat dependent on the axis ratio, with the equipartition factor increasing as $B/B_{\rm eq} \propto A^{0.21\pm0.02}$. However, the axis ratio should be correct to within a factor of a few leading to only a modest error in the magnetic field strength.
The feedback energetics measured for shock heating in FR-IIs (using dynamical models) is therefore quite robust to the surface brightness sensitivity limit. We note, however, that the efficiency with which jet energy can couple to the gas will depend on the details of the energy injection \citep{Omma+2004, Mukherjee+2016}.

The dependences of the jet power and source age on the measured source size are difficult to disentangle for FR-Is. Instead, the total energy must be directly related to the source size as $E \propto D^{3 - \beta - \xi}$ as derived from the subsonic limit of the \citet{Turner+2015} dynamical model. The energy input by the jet therefore depends on the measured source size as $E \propto D^{2}$ for typical FR-I environments with $\beta + \xi \sim 1$. The measured feedback energy is thus related to the surface brightness sensitivity limit (see Section \ref{sec:Visible linear size of flaring jet FR-Is}) as $E \propto S^{(\beta + \xi - 3)/(\beta + \xi + 1)}$, or $E \propto S^{-1}$ for typical environments. The size of 3C sources can plausibly be underestimated by at least factor of ten (see Figure \ref{fig:3Clengths}), corresponding to a factor of 100 underestimate in the feedback energy input directly from the source. Hence, it is possible that FR-Is impart feedback on significantly larger scales than inferred from their small observable sizes \citep[see also][]{Shabala+2017}.

\section{CONCLUSIONS}
\label{sec:CONCLUSIONS}

We presented a new synchrotron emissivity model that can be {applied to mock radio lobes from} existing hydrodynamical simulations and analytical models. 
The resolved emissivity model considers: (1) radiative loss processes as a function of position, (2) lobe dynamics in an arbitrary atmosphere, and (3) an analytical framework for dynamics and synchrotron emission. 
{The large-scale dynamics of the radio lobe is modelled in this work following the method of \citet{Turner+2015}. Meanwhile, the bulk flow and dispersion of the synchrotron-emitting electrons is tracked through the lobes using two-dimensional hydrodynamical simulations with a large number of tracer fields, with simulation results incorporated in the analytical model.}

\textbf{JP ageing model inappropriate for resolved FR-IIs---} {The resolved emissivity model is applied to our lobed (FR-II) and flaring jet (FR-I) dynamical models to reproduce observations of 3C436 and 3C31 respectively, to demonstrate the validity of these models. Specifically the model can reproduce: (1) the $608\rm\, MHz$ surface brightness map of 3C31, and (2) the spatial distribution of spectral ages in 3C436 inferred from X-ray observations. We find that the oldest electron ages measured by fitting the JP model to the spectra across the 3C436 radio lobes are a factor of a few younger than its dynamical age. This discrepancy seen in lobed FR-IIs is due to a violation of the JP model assumption that all electrons contributing to the SED are of similar age; their well-mixed lobes ensure electrons at least as young as half the source age are present at every point in the lobe.}

\textbf{FR-I size and FR-II width vary with sensitivity---} Limitations in measuring AGN energetics with future radio surveys were quantified using the resolved emissivity model and our dynamical models. In particular, the observed size and imparted feedback energy from FR-Is sources is dependent on the survey surface brightness detection limit, $S$, reducing as $D \propto S^{-0.5}$ and $E \propto S^{-1}$ respectively for typical environments. Meanwhile, the observed lobe widths of FR-IIs are up to ten times narrower than the physical width when viewed close to the survey detection limit.

\textbf{GPS and CSS sources are cores of FR-Is?---} The emerging FR0 class of radio sources \citep{Ghisellini+2011}, comprising gigahertz peaked and compact steep spectrum sources (GPS and CSS), may be explained by a population of low powered FR-Is. Our modelling predicts that, for uncollimated jets close to the detection limit, the surface brightness of the radio emission decreases rapidly with distance. Moreover, the radio SED is dominated by the luminosity from the younger electrons near the core acceleration site; a high-frequency turnover will not be detected in the spectra of FR-Is, with the host environment instead affecting the lower frequencies through free-free and synchrotron self-absorption.

Low powered FR-I and FR0 type objects are expected to comprise the vast majority of radio sources identified in next-generation surveys such as ASKAP EMU and MeerKAT MIGHTEE. Developing a greater understanding of the observational limitations encountered when measuring the properties of these objects in particular is therefore crucial. The analytical model presented in this work is intended as a step in this direction; {we have shown it can accurately determine the energetics of large numbers of radio AGN using an (mostly) analytical model.}

\subparagraph{}
R.T. thanks the University of Tasmania for an Elite Research Scholarship and the CSIRO for a CASS studentship. J.R. thanks the University of Tasmania for an Australian Postgraduate Award. S.S. and M.K. thank the Australian Research Council for an Early Career Fellowship, DE130101399. M.K. further thanks the University of Tasmania for a Visiting Fellowship. We would like to acknowledge the use of the high performance computing facilities provided by the Tasmanian Partnership for Advanced Computing (TPAC) funded and hosted by the University of Tasmania.

\begin{appendix}
\section{Synchrotron-emitting electron energy spectrum}
\label{sec:appendix}

{We derive here an expression for the spectral energy distribution arising from a packet of synchrotron-emitting electrons undergoing adiabatic expansion with arbitrary pressure evolution.
The radio emissivity is initially calculated analytically under the assumption that only the adiabatic expansion of the electron packet affects the evolution of the source brightness.}
The emissivity of the lobe per unit volume can be calculated \citet{Longair+2010} as

\begin{equation}
J(E) = \frac{\kappa(s) \sqrt{3\pi} B e^3 N(E) E}{16 \pi^2 \epsilon_0 c m_e (s + 1)} ,
\end{equation}

where $B$ is the lobe magnetic field strength at the time of observation, $N(E)dE$ is the number density with energies between $E$ and $E + dE$, $c$ is the speed of light, $e$ and $m_{\rm e}$ are the electron charge and mass, $\epsilon_0$ is the vacuum permittivity, and $\kappa$ is a constant dependent solely on the injection index $s$ through

\begin{equation}
\kappa(s) = \frac{\Gamma(\tfrac{s}{4} + \tfrac{19}{12}) \Gamma(\tfrac{s}{4} - \tfrac{1}{12}) \Gamma(\tfrac{s}{4} + \tfrac{5}{12})}{\Gamma(\tfrac{s}{4} + \tfrac{7}{12})} .
\label{kappas}
\end{equation}

The energy of the electrons $E$ emitting at the rest-frame observing frequency $\nu$ is found through the relation

\begin{equation}
E = \gamma m_{\rm e} c^2 = \left(\frac{2\pi \nu m_{\rm e}^{\,3} c^4}{3 e B} \right)^{1/2} .
\label{gamma}
\end{equation}

The lobe pressure is related to the electron energy density $u_{\rm e}$, thermal energy density in the particles $u_{\rm T}$ and the magnetic field energy density $u_{\rm B}$ as $p = (\Gamma_{\rm c} - 1)(u_{\rm e} + u_{\rm B} + u_{\rm T})$. The adiabatic index of the lobe plasma, $\Gamma_{\rm c}$, is generally assumed to be non-relativistic \citet[e.g.][]{KA+1997} though \citet{HK+2013} propose a relativistic lobe material. If we assume there is no energy in thermal particles \citep[i.e. $u_{\rm T} = 0$;][]{KDA+1997} and introduce $q$ as the ratio of the energy density in the magnetic field to the sum of the energies in the particles, then the electron energy density is given by

\begin{equation}
u_{\rm e} =  \frac{p}{(\Gamma_{\rm c} - 1)(q + 1)} .
\end{equation}

The magnetic field strength is similarly related to the magnetic energy density as $B = \sqrt{2\mu_0 u_{\rm B}}$ for vacuum permeability $\mu_0$, and hence is related to the pressure as

\begin{equation}
B = \left(\frac{2\mu_0 p}{\Gamma_{\rm c} - 1} \frac{q}{q + 1} \right)^{1/2} .
\end{equation}

The electron energy distribution at the time of injection follows a power-law with exponent $s$ and normalisation $N_0$, i.e. $N(E_{\rm i}, t_{\rm i})\, dE_{\rm i} = N_0 {E_{\rm i}}^{-s} dE_{\rm i}$. In the lossless case this distribution remains unchanged through to the time of observation, i.e. $N(E)\, dE = N_0 E^{-s} dE$. 
The normalisation constant $N_0$ is calculated from the electron energy density $u_{\rm e}$ and the kinetic energy of the injected electrons $E_{\rm kin}$ (i.e. excluding their rest mass) through

\begin{equation}
\begin{split}
N_0 &= u_{\rm e}(t_{\rm i}) \left(\int_{E_{\rm min}}^{E_{\rm max}} E_{\rm kin} \, {E_{\rm i}}^{-s} dE_{\rm i} \right)^{-1} \\
&= \frac{u_{\rm e}(t_{\rm i})}{(m_{\rm e} c^2)^{2 - s}} \left[\frac{\gamma_{\rm min}^{2 - s} - \gamma_{\rm max}^{2 - s}}{s - 2} - \frac{\gamma_{\rm min}^{1 - s} - \gamma_{\rm max}^{1 - s}}{s - 1}\right]^{-1} ,
\label{energydensity}
\end{split}
\end{equation}

where $E_{\rm min}$ and $E_{\rm max}$ are the low- and high-energy cut-offs to the injection electron energy distribution, and $\gamma_{\rm min}$ and $\gamma_{\rm max}$ are the corresponding Lorentz factors. These high and low-energy cut-offs are for the energy distribution of the electrons once they have reached the uniform pressure lobes. However, the electron energy distribution is more readily measured in the hotspots \citep[e.g.][]{Godfrey+2009}; we therefore derive a conversion between the hotspot and lobe energy distributions. The change in the internal energy of a packet of synchrotron-emitting electrons as it expands adiabatically from the hotspot pressure to the lower pressure lobes is given by \citet[][Equation 12]{KDA+1997}. The adiabatic expansion work done by the fluid increases the volume at the expense of reducing the energy in the synchrotron-emitting electrons. Equating this change in energy with the electron energy density in our Equation \ref{energydensity} we find the low-energy cut-off of electrons in the lobe is related to the hotspot value (subscript $h$) as

\begin{equation}
\frac{E_{\rm min}}{E_{\rm min,h}} = \frac{\gamma_{\rm min}}{\gamma_{\rm min,h}} \approx \left(\frac{p_{\rm h}(t_{\rm i})}{p(t_{\rm i})} \right)^{(1- \Gamma_{\rm c})/\Gamma_{\rm c}(s - 2)} .
\end{equation}

\citet{Kaiser+2000} proposed an empirical relationship between the hotspot and lobe pressure based on the hydrodynamical simulations of \citet{Kaiser+1999}, $p_{\rm h}(t_{\rm i})/p(t_{\rm i}) = (2.14 - 0.52\beta) (A/2)^{2.04 - 0.25\beta}$, where the density profile is approximated locally as $\rho \propto r^{-\beta}$. For a typical radio source with an axis ratio (i.e. one-sided lobe length divided by radius of cross-section) of $A = 4$, initial injection index $s = 2.4$ and expanding into a gas density profile $\rho \propto r^{-1.5}$, we find the low-energy cut-off to the electron energy distribution is a factor of 2.5 lower than at the hotspot assuming $\Gamma_{\rm c} = 4/3$, and a factor of 4.3 lower for $\Gamma_{\rm c} = 5/3$.

The electron energy spectrum evolves from the moment of injection through to the time of observation as the electron population loses internal energy. The rate of change of the Lorentz factor for the electron population is given by 

\begin{equation}
\frac{d\gamma}{dt} = \frac{a_{\rm p}}{3\Gamma_{\rm c}} \frac{\gamma}{t} - \frac{4}{3}\frac{\sigma_{\rm T}}{m_{\rm e} c} \gamma^2 (u_{\rm B} + u_{\rm C}) ,
\end{equation}

where $\sigma_{\rm T}$ is the electron scattering cross-section. The first term on the right-hand side accounts for adiabatic expansion losses of a packet of electrons which expands adiabatically as $dV_{\rm packet} \propto t^{-a_{\rm p}/\Gamma_{\rm c}}$, whilst the second term combines the loss rate from both synchrotron radiation and inverse-Compton scattering of cosmic microwave background radiation.

The distribution of electron energies at time $t$ is given by \citet{KDA+1997}

\begin{equation}
\begin{split}
N(\gamma, t, t_{\rm i})\, d\gamma &= N_0 \frac{{\gamma_{\rm i}}^{2 - s}}{\gamma^2} \left(\frac{t}{t_{\rm i}} \right)^{4a_{\rm p}/3\Gamma_{\rm c}} d\gamma \\
&= N_0 \frac{{\gamma_{\rm i}}^{2 - s}}{\gamma^2} \left(\frac{p(t)}{p(t_{\rm i})} \right)^{4/3\Gamma_{\rm c}} d\gamma .
\end{split}
\label{ngamma}
\end{equation}

The second equality is added in order to model emission from lobes with arbitrary expansion rates, described by evolution of pressure $p(t)$. We generalise the analysis of \citet{KDA+1997} to treat arbitrary lobe expansion. The injection Lorentz factor $\gamma_{\rm i}$ of electrons emitting at the observing frequency $\gamma(\nu)$ is obtained by modifying their Equation 10 to yield the following recursive relation:

\begin{equation}
\gamma_{\rm n} = \frac{\gamma_{\rm n - 1} {t_{\rm n}^{a_{\rm p}(t_{\rm n - 1}, t_{\rm n})/3\Gamma_{\rm c}}}}{t_{\rm n - 1}^{a_{\rm p}(t_{\rm n - 1}, t_{\rm n})/3\Gamma_{\rm c}} - a_2(t_{\rm n}, t_{\rm n - 1}) \gamma_{\rm n - 1}} ,
\label{injectionlorentz}
\end{equation}

where the time at the $n$-th time-step $t_{\rm n}$ and the corresponding Lorentz factor $\gamma_{\rm n}$ decrease from $(t_0, \gamma_0) = (t, \gamma)$ through to $(t_{\rm N}, \gamma_{\rm N}) = (t_{\rm i}, \gamma_{\rm i})$. Electrons injected into the lobes at very early times will suffer strong losses, and as a result not appear in the observable range of Lorentz factors at time $t$. These electrons are determined to have infinite injection Lorentz factors using Equation \ref{injectionlorentz} and will not contribute to the emissivity of the radio lobe. Finally, the constant $a_2(t_{\rm n}, t_{\rm n - 1})$ in this equation is defined in each time-step through the relation

\begin{equation}
\begin{split}
a_2(t_{\rm n}, t_{\rm n - 1}) = \frac{4 \sigma_{\rm T}}{3 m_{\rm e} c} \left[\frac{u_{\rm B}(t_{\rm n})}{a_3} {t_{\rm n}}^{-a_{\rm p}} (t_{\rm n - 1}^{\:\,a_3} - {t_{\rm n}}^{\!a_3}) + \frac{u_{\rm C}}{a_4} (t_{\rm n - 1}^{\:\,a_4} - {t_{\rm n}}^{\!a_4}) \right] ,
\label{a2}
\end{split}
\end{equation}

where $a_3 = 1 + a_{\rm p}(1 + 1/3\Gamma_{\rm c})$, $a_4 = 1 + a_{\rm p}/3\Gamma_{\rm c}$ and $a_{\rm p}$ is the exponent with which the lobe pressure changes locally, i.e. $p \propto t^{a_{\rm p}}$. Through this method we enable $a_{\rm p}$ to be parametrised as a function of time, which allows the emissivity to be modelled for generalised environments and lobe expansion.

\end{appendix}

\end{document}